\documentclass[tradiabstract]{aa}  
\usepackage{graphicx}
\usepackage{txfonts}
\usepackage{hyperref}
\usepackage{array}
\usepackage{multirow}
\usepackage{amsbsy} 
\newcolumntype{P}[1]{>{\centering\arraybackslash}p{#1}}
\newcolumntype{M}[1]{>{\centering\arraybackslash}m{#1}}
\newcolumntype{N}{@{}m{0pt}@{}}

\graphicspath{{figures/}}
\hypersetup{ bookmarks=true, unicode=true, pdftoolbar=true, pdfmenubar=true, pdffitwindow=true, pdfstartview={FitH}, pdfauthor={SJ}, colorlinks=true, linkcolor=blue,citecolor=blue, urlcolor=black, breaklinks}

\begin{document}

\title{Simple halo model formalism for the cosmic infrared background and its correlation with the thermal Sunyaev-Zel'dovich effect}

\author{A. Maniyar
        \inst{1, 2}
        \and
        M. B\'ethermin
        \inst{1}
        \and
        G. Lagache
        \inst{1}
        }

\institute{ $^1$Aix Marseille Univ, CNRS, CNES, LAM, Marseille, France\\
        $^2$Center for Cosmology and Particle Physics, Department of Physics, New York University, New York, NY, 10003, USA\\
        \email{\href{mailto:abhishek.maniyar@nyu.edu}{\textrm{abhishek.maniyar@nyu.edu}}}
        \label{inst2}
}

\date{Received 30 June 2020; accepted 16 October 2020}

\abstract
{Modelling the anisotropies in the cosmic infrared background (CIB) on all the scales is a challenging task because the nature of the galaxy evolution is complex and too many parameters are therefore often required to fit the observational data. We present a new halo model for the anisotropies of the CIB using only four parameters. Our model connects the mass accretion on the dark matter haloes to the star formation rate (SFR). Despite its relative simplicity, it is able to fit both the \textit{Planck} and \textit{Herschel} CIB power spectra and is consistent with the external constraints for the obscured star formation history derived from infrared deep surveys used as priors for the fit. Using this model, we find that the halo mass with the maximum efficiency for converting the accreted baryons into stars is $\log_{10}M_\mathrm{max} = {12.94}^{+0.02}_{-0.02} \: M_\odot$, consistent with other studies. Accounting for the mass loss through stellar evolution, we find for an intermediate-age galaxy that the star formation efficiency defined as $M_\star(z)/M_b(z)$ is equal to 0.19 and 0.21 at redshift 0.1 and 2, respectively, which agrees well with the values obtained by previous studies. A CIB model is used for the first time to simultaneously fit \textit{Planck} and \textit{Herschel} CIB power spectra. The high angular resolution of \textit{Herschel} allows us to reach very small scales, making it possible to constrain the shot noise and the one-halo term separately, which is difficult to do using the \textit{Planck} data alone. However, we find that large angular scale \textit{Planck} and \textit{Herschel} data are not fully compatible with the small-scale \textit{Herschel} data (for $\ell>3000$). The CIB is expected to be correlated with the thermal Sunyaev-Zel'dovich (tSZ) signal of galaxy clusters. Using this halo model for the CIB and a halo model for the tSZ with a single parameter, we also provide a consistent framework for calculating the CIB$\times$tSZ cross correlation, which requires no additional parameter. To a certain extent, the CIB at high frequencies traces galaxies at low redshifts that reside in the clusters contributing to the tSZ, giving rise to the one-halo term of this correlation, while the two-halo term comes from the overlap in the redshift distribution of the tSZ clusters and CIB galaxies. The CIB$\times$tSZ correlation is thus found to be higher when inferred with a combination of two widely spaced frequency channels (e.g. 143x857 GHz). We also find that even at $\ell\sim2000$, the two-halo term of this correlation is still comparable to the one-halo term and has to be accounted for in the total cross-correlation. The CIB, tSZ, and CIB$\times$tSZ act as foregrounds when the kinematic SZ (kSZ) power spectrum is measured from the cosmic microwave background power spectrum and need to be removed. Because of its simplistic nature and the low number of parameters, the halo model formalism presented here for these foregrounds is quite useful for such an analysis to measure the kSZ power spectrum accurately.}


\authorrunning{Maniyar et al.}
\titlerunning{Halo models for the CIB, tSZ and CIB-tSZ correlation}
\maketitle


\section{Introduction}\label{sec:1}

The cosmic infrared background (CIB) is made up of the cumulative emission of the infrared radiation from the dusty star-forming  galaxies throughout the Universe. It traces the star formation history of the Universe, which spans a wide range of redshift $0 \leq z \sim 6$. Measurements of the CIB can thus be used as a powerful tool to map the star formation at high redshifts \citep{Knox_2001}. Although the CIB was first detected by \cite{Puget_1996}, \cite{Lagache_2000} and \cite{Matsuhara_2000} were the first to detect and discuss the anisotropies in the CIB that are due to unresolved extra-galactic sources. The CIB includes correlated anisotropies that are excellent probes of the large-scale structure of the Universe \citep[e.g.][]{Hanson_2013}. These were first discovered by \textit{Spitzer} \citep{Lagache_2007} and were then subsequently accurately measured by \textit{Planck} and \textit{Herschel} \citep{Planck_cib_2014, Viero_2013}.  \\
Another such tracer of the underlying dark matter distribution are massive galaxy clusters. Hot electrons in these galaxy clusters Compton scatter the CMB photons and give rise to the so-called thermal Sunyaev-Zel'dovich (tSZ) effect. A part of the CIB originates in the dusty star-forming galaxies residing in the galaxy clusters. Thus, the tSZ and the CIB are expected to be correlated to a certain extent. This correlation has indeed been indirectly measured and shown to be positive by \cite{Reichardt_2012}, \cite{Dunkley_2013}, \cite{George_2015}, \cite{Reichardt_2020}, and \{\cite{Choi_2020} and thus has to be considered in the CMB power spectrum data analysis. \cite{Planck_cib_tsz_2016} also reported a measurement of the cross-correlation between the tSZ and CIB using \textit{Planck} data. In order to model the CIB$\times$tSZ signal, we in turn need accurate models of the CIB anisotropies and the tSZ. 

On large angular scales, we can use the fact that the clustering of the CIB traces the large-scale distribution of matter in the Universe up to some bias factor. This makes modelling the CIB anisotropies on large angular scales quite straightforward, as reported in for example \citet{Planck_cib_2014} and \citet{Maniyar_2018}. However, in order to describe the anisotropies on both the large and small angular scales coherently, a 'halo model' approach as developed by \cite{Cooray_2002} is generally used. With the assumption inside the halo model that all the galaxies reside in the dark matter haloes, the clustering can be considered as the sum of two components: one-halo term ($P_\mathrm{1h}$), which takes the small-scale clustering due to the correlations between the galaxies within the same halo into account; and two-halo term ($P_\mathrm{2h}$), which accounts for the clustering on large scales due to the correlations between galaxies in different haloes. Along with the assumption that all the dark matter lies within the collapsed and symmetric haloes, four ingredients are required to characterise the galaxy power spectrum within the halo modelling context: the number density of the haloes per unit mass given by the halo mass function; the halo bias between the haloes and the dark matter; the spatial distribution of the dark matter inside a halo given by the halo density profile; and the halo occupation distribution (HOD), which is a prescription for filling the dark matter haloes with galaxies. 

The first generation of the models built to interpret the CIB anisotropies was based either  on a HOD model or a combination of models of emissivities of the infrared galaxies and a linear bias \citep{Knox_2001, Lagache_2007, Amblard_2007, Viero_2009, Planck_cib_2011, Amblard_2011, Xia_2012}. These approaches assumed that the emissivity density is traced by the galaxy number density, implying that all galaxies contribute equally to the emissivity, regardless of their host halo masses. This would mean that all the galaxies have the same luminosity. However, as has been pointed out by \cite{Shang_2012}, both the luminosity and clustering of the galaxies are closely related to the host halo mass. In general, galaxies situated in more massive haloes are more luminous as a result of a higher stellar mass, and they are also more clustered. When this effect is neglected, the clustering signal on smaller angular scales might be interpreted as being due to a very high number of satellite haloes (which was the case for \citealt{Amblard_2011}) compared to what is found in numerical simulations (discussion in \citealt{Viero_2013_b}). 

Subsequently, several studies such as \cite{Shang_2012}, \cite{Viero_2013_b}, and \cite{Planck_cib_2014}  have improved upon the previous halo models by considering a link between the galaxy luminosity ($L$) and the host halo mass ($M_h$) in their model (through a $L-M_h$ relation). Although their approach is able to fit the CIB power spectra, their description of the infrared galaxies is quite simple (e.g. a single spectral energy distribution, SED, for all galaxy types, but with evolving dust temperature or without scatter on the $L-M_h$ relation). These models are useful to derive quantities such as the halo mass for the most efficient star formation, but it is hard to test their validity because a good fit can be obtained easily because of the high number of free parameters in these models. Without considering any priors, the predictions of these models for physical quantities such as the star formation rate density (SFRD) moreover do not match the corresponding constraints from the linear model or galaxy surveys \citep[e.g.][]{Planck_cib_2014}. This shows the need for physically motivated models that in addition to the power spectra can provide a  good fit or prediction for other physical quantities such as the SFRD. \cite{Bethermin_2013} used a semi-empirical model based on the observed relation between the stellar mass $M_\ast$ and the SFR, which they linked to the corresponding halo mass using abundance matching. This model gives CIB power spectra that are consistent with the measurements. Inspired by their findings of the SFR/BAR relation with respect to the halo mass (where BAR represents the baryonic accretion rate), we develop a simpler halo model for the CIB anisotropies with just four parameters. Our model connects the mass accretion onto the dark matter haloes to the corresponding SFR. 

The tSZ is measured through the so-called Compton parameter ($y$; Sec.~\ref{sec:tsz_halo}). The \textit{Planck} Collaboration provided an all-sky map of this Compton $y$ parameter and an estimate of the tSZ angular power spectrum up to $\ell \approx 1300$ \citep{Planck_tsz_2014, Planck_tsz_2016}. \cite{Boillet_2018} used these data to constrain the cosmological parameters (equation of state of the dark energy $w$ in particular) along with the tSZ parameter. We use this halo model of the tSZ to calculate the tSZ power spectra. 

\cite{Addison_2012} calculated the CIB-tSZ correlation within the halo model framework. They first presented a formalism to calculate this correlation using a CIB halo model that does not account for the dependence of the source flux on the halo mass, and then expanded their formalism to account for this effect. However, the CIB halo model they finally considered (from \cite{Xia_2012}) to calculate the CIB-tSZ correlation does not account for the aforementioned $L-M_h$ dependency. 
We present a halo model formalism to calculate the CIB$\times$tSZ cross-correlation using our new halo model for the CIB and the halo model for the tSZ from \cite{Boillet_2018}. 

This paper is structured as follows. We begin in Sect.~\ref{sec:2} by presenting our halo model for the CIB anisotropies and subsequently present the \mbox{CIBxCMB} lensing correlation within this framework. Sect.~\ref{sec:priors} then presents the constraints on the CIB model parameters through the data and corresponding results. In Sect.~\ref{sec:tsz_halo} we provide the halo model for the tSZ.
Finally, in Sect.~\ref{sec:cib-tsz-halo} we present the halo model formalism for the CIB$\times$tSZ correlation and the predictions for its power spectra and angular scale dependence\footnote{Code for calculating the CIB, tSZ, and CIB$\times$tSZ power spectra is made available online at \url{https://github.com/abhimaniyar/halomodel_cib_tsz_cibxtsz}}. Conclusions are given in Sect.~\ref{sec:concl}.  Appendices~\ref{app:sfrd}, \ref{app:moster}, and \ref{app:comparison} provide details of the CIB halo model formalism and the comparison between the \textit{Planck} and \textit{Herschel} CIB power spectrum data. \\
When required, we used a Chabrier mass function \citep{Chabrier_2003} and the Planck 2015 flat $\Lambda$CDM cosmology \citep{Planck_cosmo_2016} with $\Omega_m = 0.33$ and $H_0=67.47$\,km\,s$^{-1}$\,Mpc$^{-1}$.


\section{New halo model for the CIB power spectrum}\label{sec:2} 

The starting point for our model is the accretion of matter onto the dark matter haloes. We then connect the accretion of the baryonic gas onto the dark matter haloes to the SFR corresponding to these haloes. The SFR is defined separately for the central and satellite galaxies in a halo. Using the SFR from a given halo, we then calculate the emissivity of all the haloes of mass $M_h$ at a given redshift, which then is used to calculate the angular power spectrum of the CIB anisotropies. 

The angular power spectrum of the CIB anisotropies is defined as
\begin{equation} \label{eq:clgen}
\Big \langle \delta I_{\ell m}^{\nu} \delta I_{\ell'm'}^{\nu'} \Big \rangle = C_\ell^{\nu \times \nu'} \times \delta _{\ell \ell'}\delta_{mm'} \, ,
\end{equation}
where $\nu$ is the frequency of the observation and $I^\nu$ is the intensity of the CIB measured at that frequency. The intensity is a function of the comoving emissivity $j$ through
\begin{equation} \label{eq:inu}
\begin{split}
I^\nu & =  \int \dfrac{d\chi}{dz} a j(\nu, z) dz \\
 & =  \int \dfrac{d\chi}{dz} a \bar{j}(\nu, z) \Big( 1 + \frac{\delta j(\nu, z)}{\bar{j}(\nu, z)} \Big) dz \, ,
\end {split}
\end{equation}
where $\chi(z)$ is the comoving distance to redshift $z$, and $a = 1/(1+z)$ is the scale factor of the Universe, and $\delta j(\nu, z)$ are the emissivity fluctuations of the CIB. We expand the Eq.~\ref{eq:inu} this way between the mean value and its fluctuations because  Eq.~\ref{eq:clgen} shows that the power spectrum is calculated using the fluctuations around the mean value. Combining Eqs. \ref{eq:clgen} and \ref{eq:inu}, and using the Limber approximation \citep{Limber_1954} for the flat sky, which helps us avoid the spherical Bessel function calculations and makes the computation easier, we therefore obtain
\begin{equation} \label{eq:cl_lin2h}
C_\ell^{\nu \times \nu'} = \int \frac{dz}{\chi^2} \frac{d\chi}{dz}a^2 \bar{j}(\nu,z) \bar{j}(\nu',z)P_j^{\nu \times \nu'}(k = \ell/\chi, z) \, ,
\end{equation}
where at a given redshift, $P_j^{\nu \times \nu'}$ is the 3D power spectrum of the emissivity and is defined as
\begin{equation}
\Big \langle \delta j(k,\nu) \delta j(k',\nu') \Big \rangle = (2\pi)^3 \bar{j}(\nu)\bar{j}(\nu')P_j^{\nu \times \nu'}(k)\delta^3(k-k')\, ,
\end{equation}

Thus we have to calculate $\delta j$ to obtain the CIB angular power spectrum. For this purpose we connect (Sec.~\ref{ssec:halopower}) the SFR of the haloes with the specific emissivity $\frac{dj_\nu}{d\log M_h}(M_h, z)$ and integrate it over all the halo masses and redshift range to obtain the CIB power spectra.

\subsection{From accretion onto the dark matter haloes to SFR} \label{ssec:accr_sfr}
The dark matter haloes grow in mass over time through diffuse accretion and mergers with other lower mass haloes \citep[e.g.][]{Fakhouri_2010}.
Accretion and merger processes are also responsible for the growth of stellar mass in galaxies through galaxy-galaxy mergers and through accretion of the gas. 
The baryonic gas accreted by a given dark matter halo would form stars depending upon certain factors. This is the starting point of our model. As we mentioned earlier, previous studies used a parametric $L-M_h$ relation to derive the power spectra. Instead of assuming an $L-M_h$ relation 
with an evolution in redshift, we connect the accretion rate onto a dark matter halo described above with the corresponding SFR. This gives us an SFR$-M_h$ relation that in substance is similar to an $L-M_h$ relation. The difference between our approach and that of others is a more physical starting point of the parametrisation. 

This approach is physically motivated. This link between the accreted baryons and SFR is quite natural. The stars form out of the gas reservoirs within their host galaxies that reside in the dark matter haloes. The amount of gas at a given time depends upon the amount of gas accreted by the host dark matter halo. We assumed that this accreted gas is converted into stars with an efficiency that is a function of the mass of the halo and redshift. We used a lognormal parametrisation between the halo mass and the ratio of the SFR and the baryonic accretion rate BAR for a halo (i.e. SFR/BAR)
\begin{equation}
\label{eq:lognormal}
\frac{\mathrm{SFR}}{\mathrm{BAR}} (M_h, z) = \eta = \eta_\mathrm{max} \: e^{-\frac{\left(\log{M_h} \: - \: \log{M_\mathrm{max}}\right)^2}{2\sigma^2_{M_{h}}(z)}}
,\end{equation}
where $M_h$ is the halo mass, $M_\mathrm{max}$ represents the mass with the highest star formation efficiency $\eta_\mathrm{max}$ , and $\sigma_{M_h}(z)$ is the variance of the lognormal, which here represents the range of masses over which the star formation is efficient. SFR/BAR represents the efficiency ($\eta$) of the dark matter halo of a given mass ($M_h$) at a redshift ($z$) to convert the accreted baryonic mass into stars. 

The choice of the lognormal shape is quite logical. Several studies \citep[e.g.][]{Viero_2013, Planck_cib_2014, Maniyar_2018} have found that the dark matter haloes within the mass range $10^{12} - 10^{13}M_\odot$ form the stars most efficiently. With an empirical model, \cite{Bethermin_2013} showed in their Fig. 17 that the star formation efficiency as a function of instantaneous halo mass is highest in haloes with masses $\sim 10^{12}M_\odot$ . This mass does not change considerably over a range of redshifts, whereas the efficiency falls off drastically for masses above and below the most efficient mass. This effect can be understood physically. Below the mass for which star formation is most efficient, the gravitational potential of the dark matter halo is lower and the supernovae feedback is strong enough to remove the gas from the galaxy \citep[e.g.][]{Silk_2003} and thereby decreases the star formation. On the other side of the spectrum, at higher masses, the cooling time of the gas becomes much longer than the free-fall time \citep[e.g.][]{Keres_2005}. This suppression of the isotropic cooling of the gas could be due to the energy injection in the halo atmosphere by active galactic nuclei (AGN), which in turn suppresses the star formation \citep[e.g.][]{Somerville_2008}. Therefore we assumed a lognormal shape for the SFR efficiency whereby the star formation is highest for halo mass $M_\mathrm{max}$ , and a significant contribution to the SFR comes from a range of masses around $M_\mathrm{max}$ driven by $\sigma_{M_h}$ , and the SFR falls off considerably on very high and very low masses. 

For a given halo mass at a given redshift, the BAR is given as 
\begin{equation} \label{eq:bar}
\mathrm{BAR}(M_h, z) = \langle \dot{M}(M_h, z)\rangle \times \Omega_b(z)/\Omega_m(z) \, ,
\end{equation} 
where $\Omega_b$ and $\Omega_m$ are the dimensionless cosmological baryonic density and total matter density, respectively (thus, the ratio of the two gives the baryonic matter fraction at a given redshift, which is in fact constant with redshift because they have the same evolution with redshift). $ \dot{M}(M_h, z)$ is the mass growth rate. \cite{Fakhouri_2010} provided the mean and median mass growth rates of haloes of mass $M_h$ at redshift $z$. We used the mean mass growth rate, given as 
\begin{align}
\label{eq:fakhouri}
 {\langle \dot{M} \rangle}_\mathrm{mean} &= 46.1 \: M_\odot \: \mathrm{yr}^{-1} {\left(\frac{M_h}{10^{12}M_\odot}\right)}^{1.1} \notag \\
 & \times (1 + 1.11z) \sqrt{\Omega_m{(1+z)}^3 + \Omega_\Lambda.}
\end{align}

This approach assumes that there is no 'gas reservoir effect', that is, the accreted gas is immediately converted into stars and is not collected over time to form reservoirs in \citep{Daddi_2010} and around galaxies \citep{Cantalupo_2012}. It has been shown \citep[e.g.][]{Saintonge_2013, Bethermin_2015, Zavadsky_2015} that the depletion timescale, which is the ratio of the mass of the molecular gas to the SFR, ranges from $\sim 0.5-1$ Gyr, which is much shorter than the typical galaxy evolution timescales (several billion years). Thus, the gas reservoir effect is not expected to affect the results over the long timescale that we considered here. Along the same lines, it is also assumed that no recycled gas (i.e. the gas expelled by the supernovae) contributes to the star formation (semi-analytical models of e.g. \cite{Cousin_2015, Cousin_2019} showed that feedback from supernovae can play a part in regulating star formation). In spite of these assumptions, we show that this simple physical model describes the CIB power spectra well. 

In our model, we consider the maximum efficiency $\eta_\mathrm{max}$ and mass of maximum efficiency $M_\mathrm{max}$ to be independent of redshift, that is, they do not evolve with redshift. However, we let $\sigma_{M_h}$ evolve with redshift. The motivation for letting $\sigma_{M_h}$ evolve with redshift is that we wished to accommodate the star formation from massive haloes at higher redshifts and reduce it at lower redshifts. It has been observed \citep{Popesso_2015} that at lower redshifts ($z \leq 1.5-2$), star formation is quite inefficient in massive haloes (typical galaxy cluster environments), that is, at low redshifts, massive haloes contain mostly passive galaxies. In contrast, it has been shown that at high-redshift massive galaxies (often residing in the proto-clusters, i.e., the progenitors of the clusters at redshift zero) can have efficient star formation (e.g. \citealt{Miller_2018, Wang_2018}). Because the lognormal parametrisation leaves a tail on the high mass end, it might mimic this effect, and the choice of the this shape is therefore justified. 

Thus we let $\sigma_{M_h}$ evolve with redshift as 
\begin{equation}
\sigma_{M_h}(z) = \sigma_{M_{h0}} - \tau \times \mathrm{max}(0, z_c - z)
,\end{equation}
where $z_c$ is a redshift below which $\sigma_{M_h}$ evolves with redshift,  $\sigma_{M_{h0}}$ is the value of $\sigma_{M_h}$ above $z_c$, and $\tau$ is the parameter driving this evolution with redshift ($\tau$ here should not be confused with the optical depth parameter from the CMB analysis). Following the reasoning mentioned before, this evolution was applied only for haloes with masses greater than the mass of maximum efficiency $M_\mathrm{max}$ and below redshift $z_c$ , that is, the parametrisation is not a symmetrical lognormal below redshift $z_c$. The width of the lognormal is smaller at the side of the curve with haloes higher in mass than $M_\mathrm{max}$ below redshift $z_c$. However, above redshift $z_c$, the parametrisation is a symmetrical lognormal with no evolution in the width of the lognormal $\sigma_{M_{h0}}$. We fixed $z_c = 1.5$. Other values for $z_c$ were tried and gave approximately the same results, but the model with $z_c = 1.5$ provided the best fit for the SFRD history.

\subsection{SFR for the haloes and subhaloes} \label{ssec:sfr_halo_subhalo}
For a given value of the halo mass and redshift, we can calculate $\eta$ using Eq.~\ref{eq:lognormal} and multiply it by the corresponding BAR calculated using Eq.~\ref{eq:bar} to obtain the SFR, that is,
\begin{equation} \label{eq:SFR}
\mathrm{SFR}(M_h, z) = \eta(M_h, z) \times \mathrm{BAR}(M_h, z) \, .
\end{equation}
This is the procedure with which the SFR can be obtained for the haloes. To calculate the SFR for the subhaloes residing within these haloes, the procedure is slightly modified. We first assumed that for a given halo with mass $M_h$, the subhalo masses ($m_\mathrm{sub}$) range from $M_\mathrm{min}$ to $M_h$. In this analysis, we fixed $M_\mathrm{min}=10^5 M_\odot$. A change of the minimum mass between $10^4 M_\odot$ and $10^8 M_\odot$  changes the calculation of the power spectra only negligibly. The SFR for the subhaloes can be estimated in two ways. The first way is an approach similar to the one for the haloes, which is calculating the efficiency $\eta$ and then multiplying with the BAR value to obtain the SFR, that is, replacing $M_h$ by $m_\mathrm{sub}$ in Eq.~\ref{eq:SFR}. This assumes the same lognormal parametrisation of $\eta$ for subhaloes as of the central haloes. The other way to estimate the SFR in subhaloes is 
\begin{equation} \label{eq:sfr2}
\mathrm{SFR}_\mathrm{sub} = \mathrm{SFR}_c \times \frac{m_\mathrm{sub}}{M_h}
,\end{equation}
that is, the SFR for the subhalo is obtained by weighing the halo SFR (SFR$_c$) by the ratio of subhalo to halo mass. For every subhalo of a given halo, we estimated the SFR with both these approaches and took the smaller of the two as representative of the SFR for the subhalo. 

\begin{figure}[ht]
\centering
\includegraphics[width=9cm]{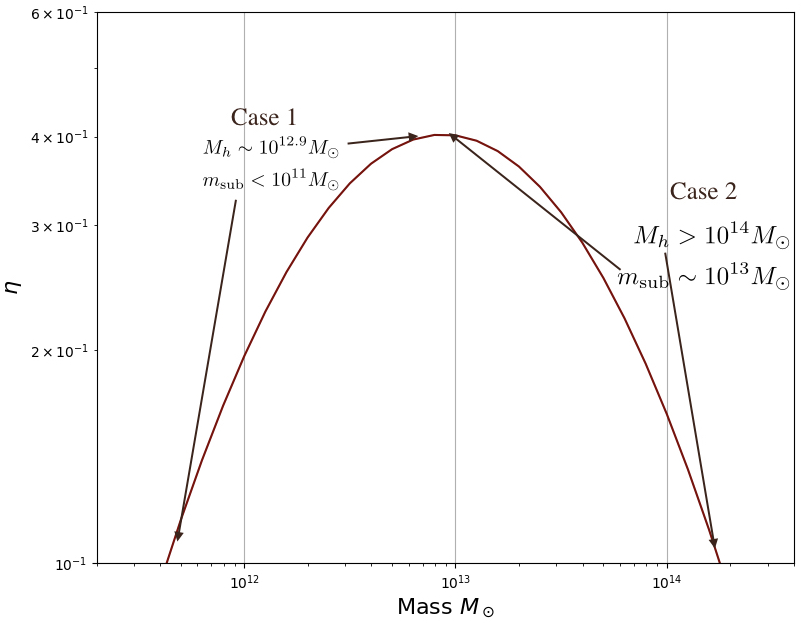}
\centering \caption{Lognormal parametrisation (Eq.~\ref{eq:lognormal}) between the halo mass ($M_\odot$) and ratio between the SFR and the baryonic accretion rate, $\eta$. We show two extreme cases: haloes near the efficiency peak  contain subhaloes with very low mass (case 1), and very massive haloes that contain subhaloes near the efficiency peak (case 2). If the same recipe were used to calculate the satellite galaxy SFR in these two cases, unphysical values might result within the assumptions of our model, and therefore we suggest two different ways to calculate the SFR for satellite galaxies (Sec.~\ref{ssec:sfr_halo_subhalo}).} 
\label{fig:lognormal_sfr}
\end{figure}

The reasoning for this is explained in Fig.~\ref{fig:lognormal_sfr}. We first consider case 1 in the figure. In this case, the main halo has a mass ($\sim 10^{12.9}M_\odot$) very near to the efficiency peak of star formation (for this particular choice of parameters for the lognormal), that is, the central galaxy forms stars very efficiently (see Eq.~\ref{eq:SFR}). This halo has subhaloes ranging from $10^5M_\odot - 10^{12.9}M_\odot$. We take the case of a subhalo with mass $< 10^{11}M_\odot$. As was pointed out before, subhaloes with very low mass have a low gravitational potential, and it is hard for them to hold on to the gas inside against the pressure from supernova feedback, for instance. Thus, they are expected to have low star formation. This is satisfied in case 1 because at lower masses, the efficiency is indeed very low and is not expected contribute significantly to the total SFR of the halo. In this case, the SFR for the subhaloes can therefore be directly estimated by substituting the halo mass ($M_h$) by subhalo mass ($m_\mathrm{sub}$) in Eq.~\ref{eq:SFR}. As an example, for a subhalo of mass $10^{11}M_\odot$ belonging to a central halo of mass $10^{12.9}M_\odot$, the SFR calculated using Eq.~\ref{eq:SFR} is 2\% of the value that we derive for the SFR obtained using Eq.~\ref{eq:sfr2}, and we therefore take the former as the SFR value. 

In the second case, the main halo is quite massive ($> 10^{14}M_\odot$) and far away from the efficiency peak. It therefore does not have strong star formation. However, in this case, this halo can contain a subhalo with a mass of about the efficiency peak ($\sim 10^{12.9}M_\odot$). According to Eq.~\ref{eq:SFR}, when we substitute $M_h$ by $m_\mathrm{sub}$ , this subhalo will have a significant amount of star formation. Again, as pointed out earlier, the gas inside massive haloes is quite hot, and there are mechanisms at play (e.g. X-ray heating and AGN feedback) that suppress the gas cooling and hence make it difficult to form stars. Moreover, the central massive galaxies in these haloes can strip the gas out of the satellite galaxies in the subhaloes and thus decrease star formation. These subhaloes are therefore not expected to contribute significantly to the SFR, in contrast to what would be obtained with Eq.~\ref{eq:SFR}. This inherently assumes instantaneous quenching, that is, the satellite galaxies are quenched at the same time as the central galaxies in a given parent halo. If we were to explicitly avoid this assumption, we would need to introduce an additional quenching parameter as a function of subhalo mass and redshift. This would result in one or two additional parameters for the model. However, as we mentioned earlier, our main purpose here is to build a very simple halo model of the CIB with as few parameters as possible. 
One way to correct for this in these cases therefore is to weight the SFR of the main halo by the mass fraction of the corresponding subhalo, that is, use Eq.~\ref{eq:sfr2} to obtain the SFR of the subhalo. The SFR in this case would be lower than the rate obtained using Eq.~\ref{eq:SFR} (substituting $M_h$ by $m_\mathrm{sub}$, of course). Again as an example, for a subhalo of mass $10^{12.9}M_\odot$ belonging to a central halo of mass $10^{14}M_\odot$, the SFR calculated using Eq.~\ref{eq:SFR} is  twice higher than the value we obtain for the SFR with Eq.~\ref{eq:sfr2} at $z = 2,$ and we therefore take the latter as the SFR value. 

Although Eq.~\ref{eq:SFR} and Eq.~\ref{eq:sfr2} seem useful to estimate the SFR for subhaloes when we have cases similar to cases 1 and 2, for every halo we therefore estimate the SFR for the corresponding subhaloes using both these methods at every redshift and select the SFR with the lower value. This automatically takes care of the extreme cases and helps us avoid adding more parameters to the model. 

\subsection{SFR to CIB power spectra}\label{ssec:halopower}
The one-halo term for the CIB power spectrum takes the clustering of the galaxies within a halo of mass $M_h$ into account and was calculated following \citet{Bethermin_2013} (where $k = \ell/\chi$),
\begin{align}
\label{eq:1halo}
C^{1h}_{\ell,\nu ,\nu'} &= \int \int \frac{d\chi}{dz}{\left(\frac{a}{\chi}\right)}^2 \Bigg[ \frac{dj_{\nu,c}}{d\log M_h} \frac{dj_{\nu',sub}}{d\log M_h}u(k, M_h, z) \notag\\
&+ \frac{dj_{\nu',c}}{d\log M_h} \frac{dj_{\nu,sub}}{d\log M_h}u(k, M_h, z) \notag\\
&+ \frac{dj_{\nu,sub}}{d\log M_h} \frac{dj_{\nu',sub}}{d\log M_h}u^2(k, M_h, z)\Bigg] \: {\left( \frac{d^2N}{d\log M_hdV} \right)}^{-1}dzd\log M_h,
\end{align}
where $\frac{d^2N}{d\log M_hdV} = \frac{dn}{d\log M_h}$ is the halo-mass function, $u(k,M_h,z)$ is the Fourier transform of the density profile describing the density distribution inside the halo (here we consider the density distribution to be a Navarro-Frenk-White (NFW) profile \citealt{Navarro_1997}), and $\frac{dj(\nu,z)}{d\log M_h}$ is the specific emissivity of the central and satellite subhaloes at a given frequency and redshift for a given halo mass as defined in \cite{Bethermin_2013}. After we calculate the specific emissivity term for the central and satellite terms, it is therefore straightforward to calculate the one-halo power spectrum. For simplicity, we omitted the $M_h$ and $z$ dependence from $\frac{dj_{\nu,c}}{d\log M_h}$ and $\frac{dj_{\nu,sub}}{d\log M_h}$ terms from all the equations. 

For the central galaxies, the differential emissivity is calculated as
\begin{align}
\frac{dj_{\nu,c}}{d\log M_h}(M_h,z) &=  \frac{d^2N}{d\log M_hdV} \times \chi^2(1+z) \times \frac{\mathrm{SFR}_c}{K} \times S^\mathrm{eff}_\nu(z),
\end{align}
where $S^{eff}_\nu(z)$ is the effective SED of the infrared galaxies at a given redshift for a given frequency. $\mathrm{SFR}_c$ is the SFR for the central galaxies with a given halo mass (Eq.~\ref{eq:SFR}). $K$ is the Kennicutt constant ($K = \mathrm{SFR}/L_\mathrm{IR}$), which has a value of $1 \times 10^{-10} M_\odot\mathrm{yr}^{-1}{\mathrm{L}_\odot}^{-1}$ for a Chabrier IMF, and $L_\mathrm{IR}$ is the infrared luminosity (8-1000$\mu$m). 

For the satellite galaxies in the subhaloes \citep{Bethermin_2013},
\begin{align}
& \frac{dj_{\nu,sub}}{d\log M_h}(M_h,z) =  \frac{d^2N}{d\log M_hdV} \times \chi^2(1+z) \, \times \notag\\
& \int \frac{dN}{d\log m_\mathrm{sub}}(m_\mathrm{sub}|M_h) \frac{\mathrm{SFR}_\mathrm{sub}}{K} \times S^{eff}_\nu(z) \times d\log m_\mathrm{sub,}
\end{align}
where $\frac{dN}{d\log m_\mathrm{sub}}$ is the subhalo mass function for the satellite galaxies with a subhalo mass $m_\mathrm{sub}$. The effective SEDs $S^{eff}_\nu(z)$ for the satellite galaxies are assumed to be the same as those of the central galaxies. The SFR$_\mathrm{sub}$ is calculated using Eqs.~\ref{eq:SFR} and \ref{eq:sfr2}, and the smaller of the two values is taken as the SFR value for those galaxies. 

In our analysis, we assumed the halo mass function from \cite{Tinker_2008} and the subhalo mass function from \cite{Tinker_2010}. $S^{eff}_\nu(z)$ are the same as we used for the linear clustering model of the CIB anisotropies from \cite{Maniyar_2018}. They were computed using the method presented in \cite{Bethermin_2013}, but assuming the new updated SEDs calibrated with \textit{Herschel} data presented in \cite{Bethermin_2015} and \cite{Bethermin_2017}. A stacking analysis was used to measure the evolution of the average mid-infrared to milimeter emission of the massive star-forming galaxies up to $z = 4$. With this technique, we found that for the main-sequence galaxies we used in the analysis, the mean intensity of the radiation field, which is strongly correlated with the dust temperature, rises with redshift. Thus the dust in these new SED templates is warmer at $z>2$ than in the previous templates used in \cite{Bethermin_2013}. We prefer these templates over the other templates (e.g. from \cite{Gispert_2000} using FIRAS measurements) because they reproduce all recent measurements of galaxy counts from the mid-IR to the radio wavelength range, including counts per redshift slice. 

The Fourier transform of the NFW profile is given as \citep[e.g.][]{Van_2013}
\begin{align} \label{eq:unfw}
u(k, M_h, z) & = \frac{3\delta_{200}}{200c^{3}} \bigg( \cos(\mu) \big[ \mathrm{Ci}(\mu + \mu c) - \mathrm{Ci}(\mu) \big] \notag \\
                   & + \sin(\mu) \big[ \mathrm{Si}(\mu + \mu c) - \mathrm{Si}(\mu) \big] - \frac{\sin(\mu c)}{\mu + \mu c} \bigg) \, ,
\end{align}
where it is to be noted that $c$ is not the speed of light; it is the so-called halo concentration parameter defined as $c (M_h, z) \equiv r_{200}(M_h, z)/r_\star(M_h, z),$ with $r_\star(M_h, z)$ being a characteristic radius \citep[see e.g.][]{Navarro_1997} and $r_{200}$ being the radius containing the mass giving 200 times the critical density of the Universe at a redshift (some studies use the average density instead of the critical density; we used the critical density), $\mathrm{Ci}(x)$ and $\mathrm{Si}(x)$ are standard cosine and sine integrals, respectively, $\mu \equiv kr_\star$ , where $\delta_{200}$ is a dimensionless amplitude, which can be expressed in terms of the halo concentration parameter as
\begin{equation}
\delta_{200} = \frac{200}{3} \frac{c^3}{\ln(1 + c) - \frac{c}{1 + c}}
.\end{equation}
The two-halo term for the power spectrum of the CIB takes the clustering between galaxies in two different haloes of mass $M_h$ and $M'_h$ into account and is calculated as \citep{Bethermin_2013} (where $k = \ell/\chi$)
\begin{align}
\label{eq:2halo}
C^{2h}_{\ell,\nu ,\nu'} &= \int \int \int \frac{d\chi}{dz}{\left(\frac{a}{\chi}\right)}^2 \Bigg[ \frac{dj_{\nu, c}}{d\log M_h} + \frac{dj_{\nu, sub}}{d\log M_h}u(k, M_h, z)\Bigg] \notag \\
& \times \Bigg[ \frac{dj_{\nu',c}}{d\log M'_h} + \frac{dj_{\nu', sub}}{d\log M'_h}u(k, M_h, z)\Bigg] \notag\\
& \times b(M_h, z) b(M'_h, z) P_\mathrm{lin}(k, z) d\log M_h \, d\log M'_h \, dz \, ,
\end{align}
where $b(M_h, z)$ is the halo bias prescription given by \cite{Tinker_2010_b}, $P_\mathrm{lin}(k, z)$ is the linear matter power spectrum, which we calculated using \texttt{CAMB}\footnote{\url{http://camb.info/}}. We used the $u(k, M_h, z)$ term only for the subhaloes. This means that in our analysis, the central galaxy is assumed to be at the centre of the halo and the satellite galaxies in the subhaloes are distributed according to the NFW profile. The clustering term given by $b(M_h, z) b(M'_h, z) P_\mathrm{lin}(k, z)$ provides the cross-power spectrum between two different haloes ($M_h$ and $M'_h$) under the assumption \citep{Cooray_2002} $P(k, M_h, M'_h) = b(M_h, z)b(M'_h, z)P_\mathrm{lin}(k, z)$. We can simplify Eq.~\ref{eq:2halo} by defining
\begin{equation}
D_\nu(k, z) = \int b(M_h, z) \Big[ \frac{dj_{\nu, c}}{d\log M_h} + \frac{dj_{\nu, sub}}{d\log M_h}u(k, M_h, z) \Big]d\log M_h
,\end{equation}
which is the emissivity of the halo weighted by the corresponding bias. Therefore Eq.~\ref{eq:2halo} becomes (where $k = \ell/\chi$)
\begin{equation}
C^{2h}_{\ell,\nu ,\nu'} = \int \frac{d\chi}{dz}{\left(\frac{a}{\chi}\right)}^2 D_\nu(k, z) D'_\nu(k, z) P_\mathrm{lin}(k, z) dz
.\end{equation}
This way of calculating the two-halo term significantly reduces the time because it reduces the number of integrals. It is also useful when the CIB-CMB lensing cross-correlation is calclutated (see Sec.~\ref{sec:cib-cmblens}). Finally, we have
\begin{equation}
C^\mathrm{CIB, clustered}_{\ell,\nu ,\nu'} = C^{1h}_{\ell,\nu ,\nu'} + C^{2h}_{\ell,\nu ,\nu'} \, ,
\end{equation}
and
\begin{equation}
C^\mathrm{CIB, tot}_{\ell,\nu ,\nu'} = C^{1h}_{\ell,\nu ,\nu'} + C^{2h}_{\ell,\nu ,\nu'} + C^\mathrm{shot}_{\ell,\nu ,\nu'} \, .
\end{equation}
Here $C^\mathrm{CIB, clustered}_{\ell,\nu ,\nu'}$ and $C^\mathrm{CIB, tot}_{\ell,\nu ,\nu'}$ describe the clustered and total CIB anisotropy power spectrum, respectively. $C^\mathrm{shot}_{\ell,\nu ,\nu'}$ is the shot-noise component, which we cannot predict using our halo model formalism here. The shot-noise component is scale independent and thus has a constant flat power spectrum for a given frequency combination. We thus fit for this constant directly at every frequency channel as described in Sec.~\ref{subsec:priors}.
 
\subsection{$\rho_\mathrm{SFR}$}
When the formalism for calculating the CIB angular power spectrum in the halo modelling context is defined, it is desirable that the halo model is able to fit or reproduce statistical properties of dusty star-forming galaxies, such as the infrared (IR) SFRD of the Universe. In the context of our halo model, the SFRD for the central galaxies is calculated as
\begin{equation} \label{eq:halosfrd}
\mathrm{SFRD}_c(z) = \int \frac{d^2N}{d\log M_hdV} \times \mathrm{SFR} (M_h, z) \times d\log M_h
,\end{equation}
and for the satellite galaxies in the subhaloes, it reads 
\begin{align}\label{eq:halosfrdsub}
\mathrm{SFRD}_\mathrm{sub} (z) &= \int \frac{d^2N}{d\log M_hdV} \Big( \int \frac{dN}{d\log m_\mathrm{sub}} \times \mathrm{SFR} (m_\mathrm{sub}, z) \notag\\
& \times d\log m_\mathrm{sub}\Big) \times d\log M_h \, ,
\end{align}
where $\mathrm{SFR} (m_\mathrm{sub}, z)$ is the satellite galaxy SFR calculated as explained in Sec.~\ref{ssec:sfr_halo_subhalo}. \\
Thus, the total SFRD is calculated by adding Eq.~\ref{eq:halosfrd} and \ref{eq:halosfrdsub} as
\begin{equation}
\mathrm{SFRD} (z) = \mathrm{SFRD}_c (z) + \mathrm{SFRD}_\mathrm{sub} (z)
.\end{equation}

\subsection{CIB-CMB lensing cross-correlation} \label{sec:cib-cmblens}
The large-scale distribution of the matter in the Universe gravitationally deflects the CMB photons that freely propagate toward us from the last scattering surface. This phenomenon is called gravitational lensing, and it leaves imprints on the temperature and polarisation anisotropies of the CMB. Because the CIB is an excellent tracer of the large-scale structure of the Universe \citep{Maniyar_2019}, its anisotropies are expected to be strongly correlated with the CMB lensing, which has indeed been observed and measured \citep{Planck_ciblensing_2014}. 
Within the context of our halo model, we can calculate this CIB-CMB lensing cross-correlation as \citep{Bethermin_2013} (where $k = \ell/\chi$)
\begin{equation} \label{eq:halolens}
C_\ell^{\nu\phi} = \int \frac{d\chi}{dz} \left(\frac{a}{\chi}\right) D_\nu(k, z) \Phi(\ell, z) P_\mathrm{lin}(k, z) dz \,,
\end{equation}
where $\Phi(\ell, z)$ is given by \citep{Challinor_2005}
\begin{equation}
\Phi(\ell, z) = \frac{3}{\ell^2} \Omega_m \Big(\frac{H_0}{c}\Big)^2 \frac{\chi}{a} \bigg(\dfrac{\chi_\star - \chi}{\chi_\star \chi}\bigg) \, ,
\end{equation}
In this equation, $\chi_\star$ is the comoving distance to the last scattering surface, $\Omega_m$ is the matter density parameter, $H_0$ is the value of the Hubble parameter today, and $a$ is the scale factor of the Universe. The measurement of this cross-correlation is not used in our likelihood to fit the CIB model parameters, but we show that our best-fit model accurately predicts this cross-correlation.

\section{Constraints on the model through data} \label{sec:priors}

\subsection{Observational constraints on the power spectra} \label{subsec:priors}
We used the CIB angular power spectra as measured by \cite{Planck_cib_2014}. The measurements were obtained by cleaning the CMB and Galactic dust from the \textit{Planck} frequency maps. They were further corrected for the SZ and spurious CIB contamination induced by the CMB template, as discussed in \cite{Planck_cib_2014}. We used the measurements at the four highest frequencies (217, 353, 545, and 857\,GHz from the HFI instrument). The CIB$\times$CIB power spectra were corrected for absolute calibration difference between the PR1 and PR2 data release of \textit{Planck}. Absolute calibration uncertainties for PR2 are equal to 6.1\% and 6.4\% at 545 and 857\,GHz (5\% of which comes from planet models), respectively \citep{Planck_PR2}. However, it has been shown that the planet calibration agrees with CMB calibrations within 1.5\% for 545\,GHz \citep{Planck_cal_2016}. 

We also used the CIB power spectra measurement from \textit{Herschel}/SPIRE data at 600, 857, and 1200\,GHz provided by \cite{Viero_2013}. \cite{Bertincourt_2016} calculated the cross-calibration factors between the power spectra measured by \cite{Viero_2013} at 600 and 857\,GHz and the power spectra measured by \cite{Planck_cib_2014} at 545 and 857 GHz. They are $1.047 \pm 0.0069$ and $1.003 \pm 0.0080,$ respectively (for the \textit{Planck} PR2 data release). Absolute calibration uncertainties for these SPIRE data are 9.5\%. 

\textit{Planck} and \textit{Herschel} measurements are given with the $\nu I_\nu = \mathrm{constant}$ photometric convention. As a result, the power spectra computed by the model need to be colour-corrected from the CIB SEDs to this convention. These colour corrections are 1.119, 1.097, 1.068, and 0.995 at 217, 353, 545, and 857\,GHz, respectively, for \textit{Planck}  \citep{Planck_cib_2014}. They are equal to 
0.974, 0.989, and 0.988 for 600, 857, and 1200 GHz channels of \textit{Herschel}/SPIRE, respectively, using the extended relative spectral response function \citep{Lagache_2019}. The CIB power spectra were then corrected as  
\begin{equation}
\label{eq:cc}
C_{\ell,\nu,\nu '}^{\mathrm{model}}\: \mathrm{x}\: \mathrm{cc}_\nu \: \mathrm{x}\: \mathrm{cc}_{\nu'} = C_{\ell,\nu,\nu '}^{\mathrm{measured}}\,.
\end{equation}

The CIB power spectra error bars do not account for the absolute calibration uncertainties. To account for these uncertainties, \cite{Bethermin_2011} introduced a calibration factor $f_\mathrm{cal}^\nu$ for galaxy number counts. Using a similar approach here, we set a Gaussian prior on these calibration factors for every frequency channel. The \textit{Herschel} power spectra were cross-calibrated with respect to the \textit{Planck} power spectra, therefore we fixed the calibration factors at the \textit{Planck} frequencies at 1 and set Gaussian priors on calibration factors $f^\nu_\mathrm{cal}$ for 600 and 857, and 1200 GHz channels from \textit{Herschel}/SPIRE with an initial value of 1.0470, 1.0030, and 1.0000 with 1$\sigma$ error bars at 0.0069, 0.0080, and 0.0500, respectively \citep{Bertincourt_2016}. In Tab. \ref{tab:onlypla} we show our best-fit values when we fit our model to the \textit{Planck} data  alone and allowed the individual calibration parameters for all the \textit{Planck} channels to vary with a 5\% prior on them. The posterior values show that the error bars on the calibration parameters for \textit{Planck} (which are very close to 1) are indeed very small. For our purposes here, the assumption of taking a perfect calibration for \textit{Planck} therefore does not strongly affect the results.

When only \textit{Planck}/HFI data are used, it is hard to differentiate between the one-halo and the shot-noise term, and therefore, \cite{Planck_cib_2014} placed strong flat priors on the shot noise at all the frequencies.  They took values of the shot noise based on the model of \cite{Bethermin_2012}, andwas  the width of the prior given by their $1\sigma$ error bars. 
However, using \textit{Herschel}/SPIRE data, the CIB power spectra were measured to very small scales ($\ell_\mathrm{max} \sim 30360$). At these highest multipoles, the power spectra are dominated by the shot noise, and it is therefore better possible to constrain the shot noise together with the one-halo term with these data than using the \textit{Planck}/HFI data alone. 
Because \textit{Herschel}/SPIRE data can distinguish between the one-halo and the shot-noise term, this paves the way to include shot noise as a parameter for every pair of the power spectra with broad priors, that is, 10 shot-noise parameters for HFI, as done by \cite{Planck_cib_2014}, and 6 for SPIRE data. This would be 16 additional parameters. In order to circumvent this problem, we calculated the correlation of the 10 shot-noise parameters for HFI and 6 for SPIRE using the values predicted by the model of \cite{Bethermin_2012}. We then fit for the shot-noise parameters for only the auto-power spectra and used the correlation matrix to obtain the shot-noise level for every frequency pair. Thus, in the end we have only 1 shot noise parameter per frequency, that is, 4 for HFI and 3 for SPIRE. We placed broad flat priors 
on them with sufficient width ([0.1 - 2] times  the value estimated by the \cite{Bethermin_2012} model on either side) to avoid biasing the parameter estimation. We also tested the model by placing the shot noise at all the SPIRE auto- and cross-power spectra as parameters instead of only taking the auto-power as the parameters because SPIRE might be able to differentiate between the clustering and the shot noise because its angular resolution is very high. We find similar values for the shot noise in both cases and therefore selected the case with fewer parameters.

\subsection{External observational constraints}
In addition to the CIB auto- and cross-power spectra, we used external constraints from the mean CIB intensity values at different frequencies, and the SFRD measurements at different redshifts:
\begin{enumerate}
\item
As mentioned before, the previously used halo models over-predicted the SFRD of the Universe compared to the SFRD measured by external groups using the infrared luminosity functions. We would like the halo model to be able to fit the CIB power spectra and also produce the correct SFRD history. Thus, we used the $\rho_{\mathrm{SFR}}$ measurements at different redshifts that were obtained by measuring the infrared luminosity functions from \citet{Gruppioni_2013, Magnelli_2013, Marchetti_2016} as priors while performing the fit. In order to account for the different sets of cosmological parameters used by them, we converted the SFRD values into the observed flux between $8-1000 \mu m$ per redshift bin per solid angle, as done in \cite{Maniyar_2018}. 

In Appendix~\ref{app:sfrd} we show the SFRD history produced by our halo model when this external prior is not considered in the fit. There is indeed a significant change in the SFRD history predicted by the model when we do not include any priors at all. This means that including this prior is quite important to obtain physical results from our model.
\item
The mean level of the CIB was deduced at different frequencies using the galaxy number counts. We used these measurements as constraints on our model. Similar to the case of the power spectra, the mean level of the CIB computed by the model needs to be colour-corrected. The values of the mean level with their corresponding frequencies and the colour corrections we used are given in Table 2 of \cite{Maniyar_2018}. 
\end{enumerate}

\subsection{Fitting the data}
We performed a  Markov chain Monte Carlo (MCMC) analysis in the global CIB parameter space using the Python package `emcee' \citep{emcee}. We have a 14-dimensional parameter space: \newline - physical model parameters $\{\eta_\mathrm{max}, M_\mathrm{max}, \sigma_{M_{h0}}, \tau \}$, \newline - calibration factors $\{f_{600}^\mathrm{cal}, f_{857}^\mathrm{cal}, f_{1200}^\mathrm{cal} \}$, \newline - shot noises $\{\mathrm{SN^{pl}_{217}}, \mathrm{SN^{pl}_{353}}, \mathrm{SN^{pl}_{545}}, \mathrm{SN^{pl}_{857}}, \mathrm{SN^{sp}_{600}}, \mathrm{SN^{sp}_{857}}, \mathrm{SN^{sp}_{1200}} \}$. 

The global $\chi^2$ has a contribution from the CIBxCIB, priors on calibration factors, and priors imposed by the external observational constraints mentioned above. 
We assumed Gaussian uncorrelated error bars for measurement uncertainties, which simplifies the covariance matrix that ultimately contains only diagonal terms.

\subsection{Results} \label{ssec:res}
\begin{figure*}[ht]
\centering
\includegraphics[width=\textwidth]{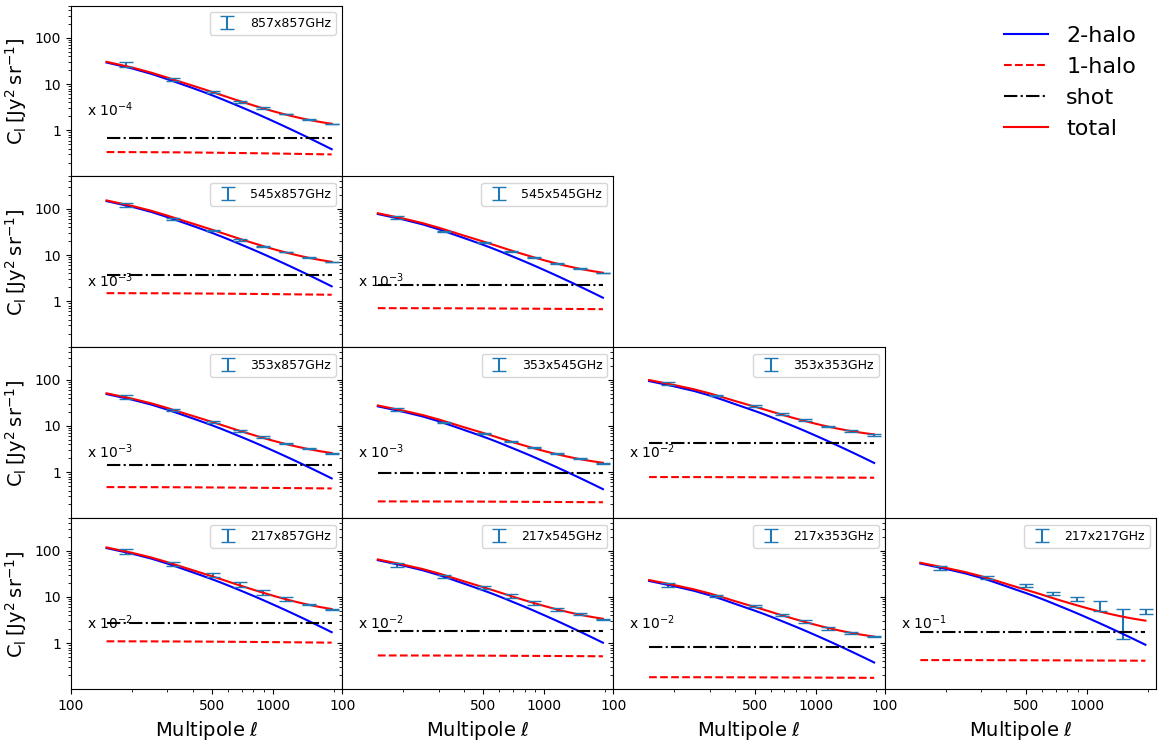}
\centering \caption{Measurements of the CIB auto- and cross-power spectra obtained by \textit{Planck}/HFI \citep{Planck_cib_2014} and the best-fit CIB halo model with its different components. }
\label{fig:cibfithalopl}
\end{figure*}

\begin{figure*}[ht]
\centering
\includegraphics[width=\textwidth]{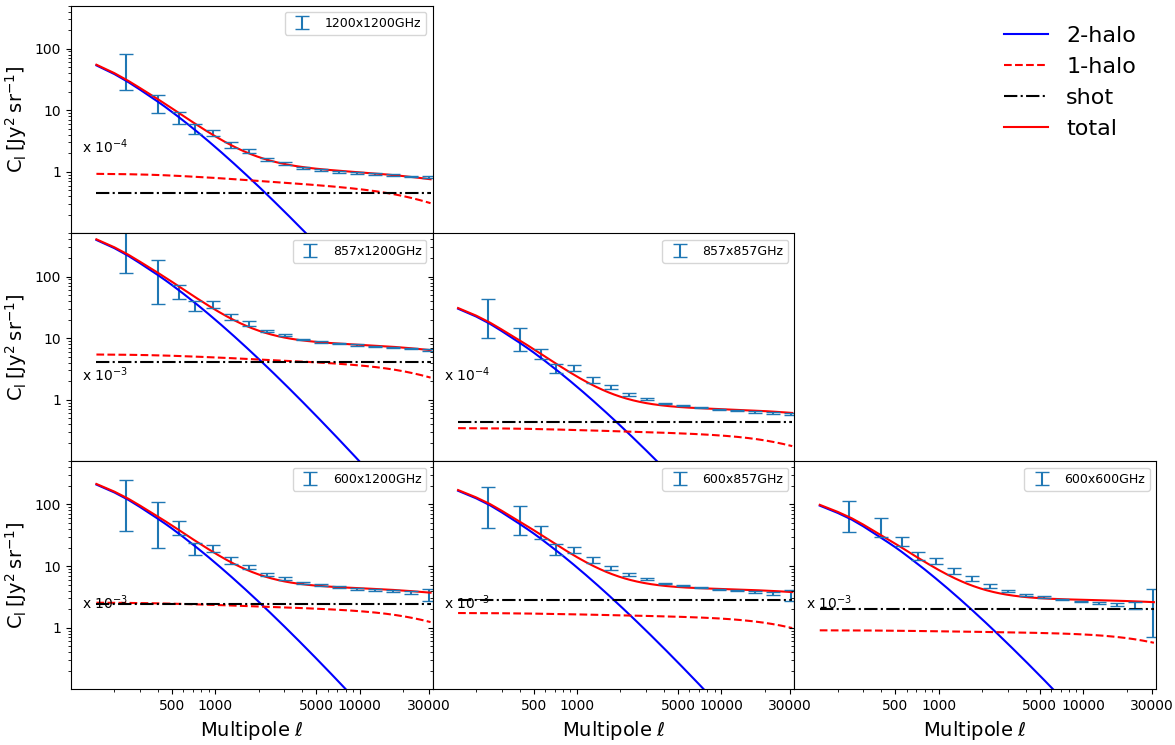}
\centering \caption{Measurements of the CIB auto- and cross-power spectra obtained by \textit{Herschel}/SPIRE \citep{Viero_2013} and the best-fit CIB halo model with its different components. }
\label{fig:cibfithalosp}
\end{figure*}

\begin{figure}[ht]
\centering
\includegraphics[width=9cm, height = 10cm]{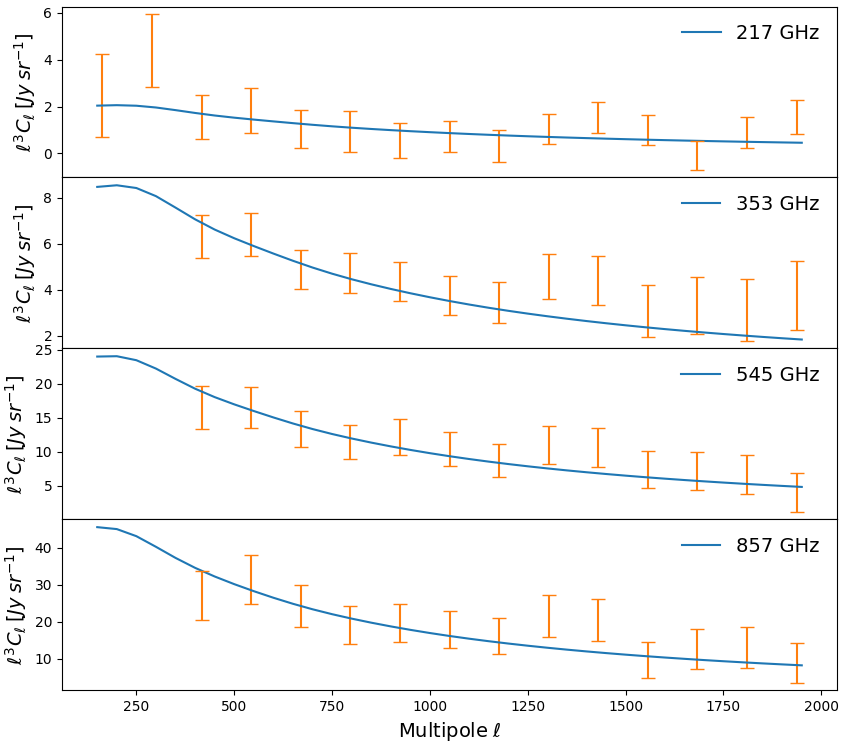}
\centering \caption{CIBxCMB\,lensing cross-power spectra measured by \cite{Planck_ciblensing_2014} in orange and our best-fit model in blue curves.}
\label{fig:lens_fit_halo}
\end{figure}

Figures~\ref{fig:cibfithalopl} and \ref{fig:cibfithalosp} show the best fit for the halo model to the observational data points for all CIBxCIB auto- and cross-power spectra from \textit{Planck}/HFI and \textit{Herschel}/SPIRE, respectively. For most of the cases, the one-halo term is smaller than the shot noise and can potentially be ignored on smaller scales. 
Based on the best-fit parameters and Eq.~\ref{eq:halolens}, we calculated the CIBxCMB\,lensing power spectra. Figure~\ref{fig:lens_fit_halo} shows these CIBxCMB\,lensing power spectra. The cross-correlation values and error bars are available for the six \textit{Planck} HFI channels (100, 143, 217, 353, 545, and 857\,GHz) and are provided in \cite{Planck_ciblensing_2014}. These values range from $\ell = 163$ to $\ell = 1937,$ and as discussed in \cite{Planck_ciblensing_2014}, the non-linear term can be neglected in this range of multipoles. As was done for the CIB, these power spectra were corrected for absolute calibration difference between the PR1 and PR2 data release of \textit{Planck}. The best-fit model agrees very well with the data points. 

We present the results from our fit for our model parameters in Table \ref{tab:3halo}. The shot-noise values provided here have to be multiplied with the corresponding colour corrections at each frequency to obtain them in the $\nu I_\nu = \mathrm{constant}$ convention. The posterior of all the parameters with a Gaussian prior (calibration factors) are within a $1\sigma$ range of the prior values. We derive a $\chi^2$ of 113 for 80 data points for \textit{Planck} power spectra. Similarly, we derive a $\chi^2$ of 247 for \textit{Herschel} power spectra for 102 data points. When we only fit for the \textit{Planck} data without the \textit{Herschel} power spectrum data but with the external priors, we derive a $\chi^2$ of 85 for \textit{Planck} for 80 data points. Similarly, fitting only for the \textit{Herschel} power spectra with the external priors results in the $\chi^2$ of 221 for 102 data points. This shows that when we individually fit for either the \textit{Planck} or \textit{Herschel} data, we derive a better $\chi^2$ value than when we fit for both of them together. 

We also performed this analysis with the updated values of the mean level of the CIB at 350, 350, and 500 $\mu$m \textit{Herschel}/SPIRE bands used as priors in our work \citep{Duivenvoorden_2020}. With the updated priors, the $\chi^2$ for \textit{Herschel} is reduced to 222, but the $\chi^2$ for \textit{Planck} data remains the same. 

In none of the approaches does the $\chi^2$ value for the \textit{Herschel} power spectra appear to fare as well as the \textit{Planck} value. 
Similarly, using a different halo model with a larger number of parameters, \cite{Viero_2013} did not obtain a good $\chi^2$ value ($\chi_\mathrm{reduced}^2 \sim 1.6$). We investigated this problem by testing the compatibility of the \textit{Planck} and \textit{Herschel} measurements. The details of the tests we performed are given in Appendix~\ref{app:comparison}. 

In Fig.~\ref{fig:cib_alldata} we show the measurements of the CIB power spectra from \textit{Planck}/HFI \citep{Planck_cib_2014}, \textit{Herschel}/SPIRE \citep{Viero_2013}, and \cite{Lenz_2019} at 857\,GHz. We also show the best-fit values of the one-halo, shot noise, and the total power spectrum when the model was fit to the \textit{Planck} or \textit{Herschel} data alone. The \textit{Herschel}/SPIRE data points and the best-fit curves were scaled to obtain them as they would be measured using the \textit{Planck} filters at 857\,GHz. The procedure for this is provided in Appendix A1 of \cite{Lagache_2019}. A similar trend is observed for the power spectra at 545\,GHz. The \textit{Herschel} and \textit{Planck} CIB measurements agree very well at large scales and up to $\ell$=2000. The extrapolation of the \textit{Planck} best fit at higher $\ell$ largely overestimates the \textit{Herschel} data points, however. This mostly comes from the shot noise, which is not compatible for \textit{Planck} and \textit{Herschel}. While the difference of flux cuts (710\,mJy for \textit{Planck} and 300\,mJy for \textit{Herschel}) would give a variation of at most 12\% of the shot-noise level\footnote{The difference in shot noise according to these two flux cuts was computed using 13 different models we had in hand.}, a factor of $\sim$1.9 is measured between the two. This discrepancy cannot be reconciled considering the difference in amplitude of the one-halo term. This discrepancy of the \textit{Herschel} and \textit{Planck} shot-noise measurements, and with the shot-noise values derived from models of galaxy number counts, has previously been pointed out by \cite{Lagache_2019}. Inconsistencies also exist inside a single experiment. For example, we show in Fig.~\ref{fig:cib_thacker_spire} the comparison of the CIB power spectra measured by \cite{Thacker_2013} and \cite{Viero_2013} for the same flux cut of 50\,mJy at 857\,GHz. The two measurements are clearly different at very high multipoles. 

We explored two ways of reconciling the \textit{Planck} and \textit{Herschel} measurements using our halo model. In the first approach, we broadened the error bars on the calibration factors for the 600, 857, and 1200 GHz \textit{Herschel} channels to 0.05 each with their central values at 1.00 instead of using the values obtained for the relative calibration between \textit{Planck} and \textit{Herschel} by \cite{Bertincourt_2016}, as done in our original approach. In this setting, we indeed find a very good fit to the \textit{Herschel} power spectra, with a $\chi^2$ of 127 when fit together with \textit{Planck} data (Tab.\,\ref{tab:hers+pla+freecal}), and 137 when the \textit{Herschel} data are fit alone (Tab\,\ref{tab:onlyhers}) compared to the $\chi^2$ of 221 in Tab\,\ref{tab:hers+freecal} for 102 data points. However, the best-fit value for the calibration factors $f^\nu_\mathrm{cal}$ as shown in Tab.~\ref{tab:hers+pla+freecal} and \ref{tab:hers+freecal} is too high. As mentioned before, the cross-calibration between \textit{Herschel}/SPIRE and \textit{Planck}/HFI has been measured very precisely at 545 and 857 GHz by \cite{Bertincourt_2016} (the 545\,GHz \textit{Planck} channel has been calibrated to better than 1.5\%, see Sec.\ref{subsec:priors}). Thus the best-fit values obtained for the $f^\nu_\mathrm{cal}$ parameters are unrealistically high and cannot be accepted. In the second approach, we keep the original priors for the \textit{Herschel} $f^\nu_\mathrm{cal}$ parameters from \cite{Bertincourt_2016} and fit the data for low multipoles only, that is, $\ell<3000$. Results are provided in Tab.~\ref{tab:hers+lowell3000} and Fig.~\ref{fig:hers+lowell3000}, which show that we obtain a good $\chi^2$ (36 for 42 data points), and the shot-noise levels for \textit{Planck} and \textit{Herschel} are quite similar (see Tab.~\ref{tab:onlypla} and \ref{tab:hers+lowell3000}), which is what we expect. This shows that the \textit{Herschel} data at $\ell>3000$ are not compatible with the values expected from the $\ell<3000$ \textit{Planck} and \textit{Herschel} data. 

\begin{table*}
 \centering
\begin{tabular}{ccccc}
\hline
Halo model parameters & $\eta_\mathrm{max}$ & $\log_{10}M_\mathrm{max}$ & $\sigma_{M_{h0}}$ & $\tau$  \\ [6pt]
 $z_c = 1.5$ (fixed) & $0.42\substack{+0.03 \\ -0.02}$ & $12.94\substack{+0.02 \\ -0.02} \: M_\odot$ & $1.75\substack{+0.12 \\ -0.13}$ & $1.17\substack{+0.09 \\ -0.09}$  \\ [6pt]
\hline
HFI shot noise & $\mathrm{SN^{pl}_{217}}$ & $\mathrm{SN^{pl}_{353}}$ & $\mathrm{SN^{pl}_{545}}$ & $\mathrm{SN^{pl}_{857}}$  \\ [6pt]
 & $13.88\substack{+0.71 \\ -0.71}$ & $353.21\substack{+9.91 \\ -8.66}$ & $2003.95\substack{+37.11 \\ -34.16}$ & $7036.95\substack{+190.48 \\ -179.71}$   \\ [6pt]
\hline
SPIRE shot noise & $\mathrm{SN^{sp}_{600}}$ &  $\mathrm{SN^{sp}_{857}}$ & $\mathrm{SN^{sp}_{1200}}$ \\ [6pt]
 & $1917.56\substack{+62.56 \\ -54.50}$ & $4273.21\substack{+164.88 \\ -148.09}$ & $4293.92\substack{+399.66 \\ 329.33}$ \\ [6pt]
\hline
HFI/SPIRE cross-calibration & $f_{600}^\mathrm{cal}$ & $f_{857}^\mathrm{cal}$ &  $f_{1200}^\mathrm{cal}$ \\ [6pt]
 & $1.06\substack{+0.01 \\ -0.01}$ & $1.02\substack{+0.01 \\ -0.01}$ & $1.04\substack{+0.03 \\ -0.03}$ \\ [6pt]
\hline
\end{tabular}
\newline
\centering \caption{Marginalised values of all the model parameters given at a 68 \% confidence level. The upper row shows the name of the parameter and the row below shows its value. We derive a $\chi^2$ value of 113 for 80 data points for \textit{Planck}, and 247 for 102 data points for \textit{Herschel}.}
\label{tab:3halo}
\end{table*}

The mass of the dark matter haloes for converting the accreted baryons into stars most efficiently is $\log_{10}M_\mathrm{max} = 12.94\substack{+0.02 \\ -0.02} \: M_\odot$. The highest efficiency mass found here is slightly on the high side of the range $\log_{10}M = 12.1\substack{+0.50 \\ -0.50}$ M$_{\odot}$ to $12.6\substack{+0.10 \\ -0.10}$ M$_{\odot}$ found by \cite{Viero_2013} and \cite{Planck_cib_2014}, but it agrees well with \cite{Chen_2016} for faint SMGs of $\log_{10}M = 12.7\substack{+0.1 \\ -0.2}$ and $\log_{10}M = 12.77\substack{+0.128 \\ -0.125}$ using the linear clustering model for the CIB anisotropies \citep{Maniyar_2018}. The error bars on the $M_\mathrm{max}$ are quite tight. Based on our best fit, we find that remaining three physical parameters of the model ($\eta_\mathrm{max}$, $\sigma_{M_{h0}}$, and $\tau$) are quite correlated amongst themselves ($\sim 90\%$), while $M_\mathrm{max}$ is much less correlated with these parameters ($\sim 10\%$). Furthermore, we did not vary $M_\mathrm{max}$ with redshift. This, combined with the fact that we have just four parameters in the model, appears to be the reason that the constraints on $M_\mathrm{max}$ are much tighter than those derived from other studies. 

With this simple model, $\eta_\mathrm{max}$ , that is, the highest efficiency with which the accreted baryonic gas forms into stars, is $\sim$40\%. We recall that we did not account for the effects such as quenching, AGN and supernovae feedback, and their evolution with redshift through explicit parameters. This efficiency parameter $\eta_\mathrm{max}$ can therefore be considered as the efficiency of converting the accreted baryons into stars after marginalising over these effects. \cite{Behroozi_2013} reported that the baryonic accretion efficiency as defined by us, that is, SFR/BAR, varied between 20-40\% across all redshifts, which is consistent with our result. Moreover, \cite{Moster_2018} calculated the star formation efficiency as $M_*(z)/M_b(z) = M_*(z)/(M_h(z) \times f_b) = M_*(z)/M_h(z) \times \Omega_m/\Omega_b,$ where $M_*(z)$ is the stellar mass in a given halo at a given redshift. They reported that this value is 0.17 at redshift $z=0.1$ and 0.2 at $z=2.0$. 

We derived a similar quantity from our model by integrating the mean SFR and BAR over cosmic times for various halo mass seeds at z=10. We find a maximum efficiency of 0.34 at $z=0.1$ and 0.37 at $z=2.0$, which is higher than the values reported by \cite{Moster_2018}. However, in this approach, we did not take the mass loss from stellar evolution into account. Similar to \cite{Zahid_2014}, and assuming 45\,\% mass loss, corresponding to what was found by \cite{Leitner_2011} for an intermediate-age galaxy, we  find $M_*(z)/M_b(z) = 0.19$ at redshift $z=0.1$ and 0.21 at $z=2$. These values agree very well with the previous values of \cite{Moster_2018}. 

Our model provides a non-zero value for $\tau,$ which means that it supports an evolution of the width of the lognormal parametrisation over time. This agrees with the observed suppression of star formation in massive haloes at low redshifts. 

We tried different parametrisations for $\eta$ instead of the lognormal. The details of one such parametrisation are provided in Appendix~\ref{app:moster}. Because of its simplicity and because it provides physical results, we continue with the lognormal parametrisation.

\begin{figure}[ht]
\centering
\includegraphics[width=9cm]{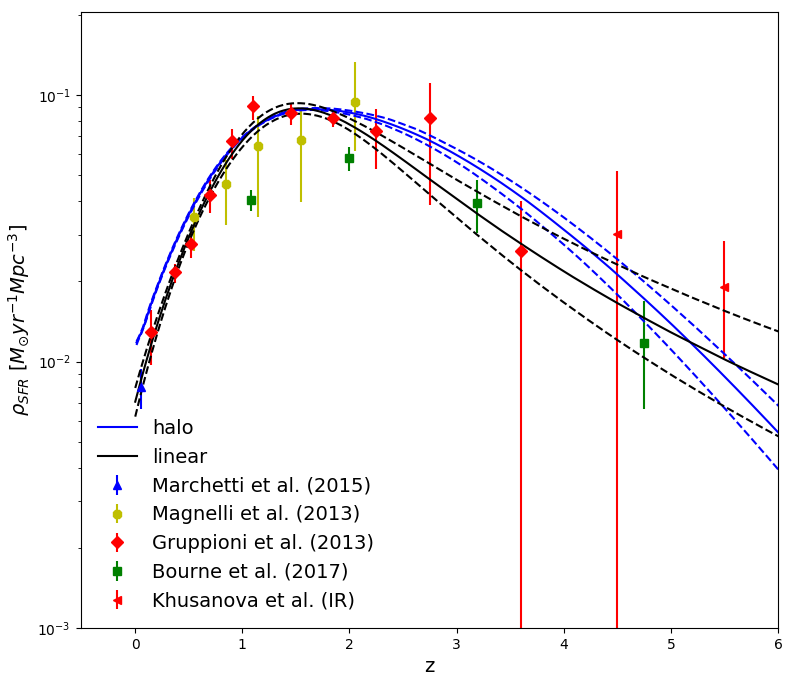}
\centering \caption{Measurements of the infrared SFRD using galaxies \citep{Madau_2014}. The solid black (blue) line shows the SFRD as constrained by the linear (halo) model with the dotted black (dotted blue) lines showing the 1$\sigma$ regions around it. We also show the SFRD value determined by Khusanova et al. completely independently using the data from the ALMA ALPINE large program (private communication).} 
\label{fig:sfrd_halo}
\end{figure}

Finally, as pointed out in the introduction, one of the shortcomings of the halo model used by \cite{Planck_cib_2014} is that it is does not match the infrared SFRD derived from galaxies. A primary reason for this is that they did not include the independent measurements of the SFRD from galaxy surveys as priors in their likelihood analysis. Consequently, the result from their linear clustering model is not consistent with the halo model either. It is therefore a good consistency check of the model to verify whether the prediction from the linear model and the infrared SFRD measurements from galaxies (which are added as priors in our likelihood) match the SFRD derived from the best fit of the model. Figure~\ref{fig:sfrd_halo} shows the SFRD constrained by our CIB halo model along with the SFRD constrained by the linear model presented in \cite{Maniyar_2018}. Although at low redshifts the SFRD constrained by the halo model is slightly higher than the measurements from galaxies or the constraints from the linear model, it is overall consistent with the latter two. Thus the SFRD obtained from the halo model is able to pass through the SFRD priors derived from galaxy surveys. This is again impressive, considering the fact that we modelled the CIB using only four parameters. 

\section{tSZ halo model} \label{sec:tsz_halo}
While at lower multipoles ($\ell < 1000$) the tSZ angular power spectrum is sensitive to the amplitude of the matter fluctuations, at higher multipoles ($\ell > 1000$) the power spectrum also depends on the details of the pressure profile of the intra-cluster gas within the haloes. This means that along with the halo mass function, the model also has to account for the pressure profile of the gas. 

The power spectrum of the tSZ is calculated as \citep[e.g.][]{Boillet_2018}
\begin{align} 
\label{eq:tsz-1h}
C_{\ell, \nu, \nu'}^{1h} &= f(\nu) f(\nu') \int \frac{dV}{dz} dz \: \times \notag \\  
& \int_{m_\mathrm{min}}^{m_\mathrm{max}} d\log M_h \frac{d^2N}{d\log M_h d\log V} |y_\ell (M_h, z)|^2,
\end{align}
where $f(\nu)$ gives the frequency dependence of the tSZ. It is given by 
\begin{equation}
f(x) = x\dfrac{e^x + 1}{e^x - 1} - 4 \: \: \mathrm{where} \: \: x = \dfrac{h_p\nu_\mathrm{obs}}{k_BT_\mathrm{CMB}} \, ,
\end{equation}
where $h_p$ is the Planck constant, and $k_B$ is the Boltzmann constant. $y_\ell (M_h, z)$ represents the two-dimensional Fourier transform of the electron pressure profile $P_e$ for a halo of mass $M_h$. It is given as \citep[e.g.][]{Komatsu_2001}
\begin{equation} \label{eq:yell_tsz}
y_\ell (M_h, z) = \frac{\sigma_T}{m_e c^2} \frac{4\pi r_{500}}{\ell_{500}^2} \int_{x_\mathrm{min}}^{x_\mathrm{max}} dx x^2 \frac{\sin(\ell x/\ell_{500})}{\ell x/\ell_{500}} P_e(x) \,.
\end{equation}
In this equation, $c$ is the speed of the light, $\sigma_T$ is the Thomson cross-section, $m_e$ is the electron mass, $x \equiv r/r_{500}$ , with r being the radial distance from the centre of the halo, $r_{500}$ is the radius of the sphere that contains the over-density mass $M_{500c}$ of 500 times the critical density of the universe, and $\ell_{500} \equiv d_A/r_{500}$ , with $d_A$ being the angular diameter distance. 

As was done for the CIB modelling, we used the NFW profile to represent the density distribution inside the dark matter halo. Using this distribution, we have
\begin{equation} \label{eq:pressure_tsz}
P_e(x) = C \times P_0 {(c_{500} x)}^{-\gamma} {[1 + {(c_{500} x)}^{\alpha} ]}^{(\gamma - \beta)/\alpha} \,
,\end{equation}
where parameters ($P_0$, $c_{500}$, $\gamma$, $\alpha$, $\beta$) were set to their best-fitting values obtained by \cite{Planck_tsz_2013}. They are equal to 6.41, 1.81, 0.31, 1.33, and 4.13, respectively. 
The coefficient $C$ goes with the mass as
\begin{equation}
C = 1.65 \: {\Bigg(\frac{h}{0.7}\Bigg)}^2 {\Bigg(\frac{H}{H_0}\Bigg)}^\frac{8}{3} {\Bigg[\frac{(h/0.7) \tilde{M}_{500c}} {3 \times 10^{14} M_\odot} \Bigg]}^{\frac{2}{3} + 0.12}  \mathrm{eV} \: \mathrm{cm}^{-3} \, ,
\end{equation}
where $H$ is the Hubble constant at redshift $z,$ with $H_0$ being its local value and $h$ the reduced Hubble constant ($h = H_0/100$). The mass used here $\tilde{M}_{500_c}$ is not necessarily the true mass, but can contain a bias due to observational effects 
and non-thermal pressure, that is, there can be a difference between the true mass of the cluster and that obtained assuming hydrostatic equilibrium. In order to account for this possible bias, a variable $B$ is often used, which relates the true mass $M_{500_c}$ to $\tilde{M}_{500_c}$ as $\tilde{M}_{500_c} = M_{500_c}/B$. In our analysis, we took the mass and redshift range to be the same as was used for the CIB model with $x_\mathrm{min} = 10^{-6}$ and $x_\mathrm{max} = 10$ 
in Eq.~\ref{eq:yell_tsz}. 

Because massive clusters are rare, the contribution of the  two-halo term to the total tSZ angular power spectrum is far lower than that of the one-halo term and can therefore be neglected \citep{Komatsu_1999} for $\ell \geq 300$. It is important only for $\ell \leq 300,$ but on these scales, the primary CMB anisotropies are clearly dominant. However, as we show in Sec.~\ref{ssec:cibxtsz}, the two-halo term plays a significant role in the two-halo term of the CIB$\times$tSZ cross-power spectra and cannot be ignored. In our analysis, we therefore do not neglect this component.


It is straightforward to calculate the two-halo component of the power spectrum as \citep[e.g.][]{Salvati_2018}
\begin{align} \label{eq:tsz-2h}
C_{\ell, \nu, \nu'}^{2h} &= f(\nu) f(\nu')  \int_{z_\mathrm{min}}^{z_\mathrm{max}} \frac{dV}{dz} dz P_\mathrm{lin}(k = \ell/\chi, z) \notag \\
& \Bigg\{ \int_{m_\mathrm{min}}^{m_\mathrm{max}} d\log M_h \frac{d^2N}{d\log M_h d\log V} b(M_h, z) y_\ell (M_h, z) \Bigg \}^2 \, .
\end{align}
The total tSZ power spectrum is the sum of the one-halo and two-halo components:
\begin{equation}
C_{\ell, \nu, \nu'}^{\mathrm{tSZ-tot}} = C_{\ell, \nu, \nu'}^{1h} + C_{\ell, \nu, \nu'}^{2h} \, .
\end{equation}

 $f(\nu)$ is negative for $\nu \leq 217$ GHz. This means that the tSZ power spectrum for certain choices of frequency combinations will be negative. The case for the CIB-tSZ correlation described in Sec.~\ref{sec:cib-tsz-halo} is similar. To calculate the $f(\nu)$ parameter correctly for a broadband filter, we have to convolve the tSZ frequency dependence with the bandpass filter at that particular frequency. The $f(\nu)$ parameter is used to convert the dimensionless $y$ parameter into CMB temperature units. Although the plots presented here for the tSZ power spectra are made independent of frequency (dimensionless) by dividing out the factors of $f(\nu)$, plots for the CIB-tSZ correlation are calculated at the value of the $f(\nu)$ at that particular frequency channel and are convolved with the filters. For 100, 143, 217, 343, 545, and 857 GHz corresponding to the \textit{Planck}/HFI frequencies, f($\nu$) is -1.51, -1.04, -0.01, 2.24, 5.60, and 11.09, respectively, when it is not convolved with the respective bandpass filters. The effective values become -4.03, -2.79, 0.19, 6.21, 14.46, and 26.34, respectively, after convolving with the \textit{Planck}/HFI bandpass filters at these respective frequencies, as given in Tab.1 of \cite{Planck_tsz_2016}. 

\begin{figure}[ht]
\centering
\includegraphics[width=9cm]{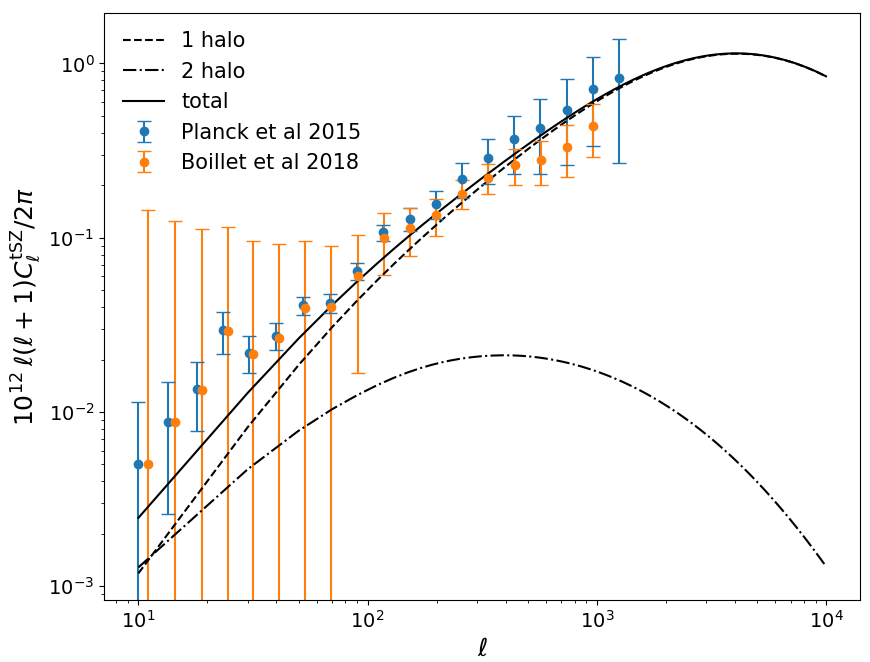}
\centering \caption{Predictions for the tSZ power spectrum based on our model for the given parameter values. The frequency dependence of the power spectrum through $f(\nu)$ has been divided out and the power spectrum is dimensionless.} 
\label{fig:tsz_f1}
\end{figure}

Because the values of the parameters ($P_0$, $c_{500}$, $\gamma$, $\alpha$, $\beta$) are not varied, this allows us to tabulate $y_\ell$ , which speeds up the process considerably. Figure \ref{fig:tsz_f1} shows the prediction for the tSZ auto-power spectrum with this model and with the given values of the parameters along with B=1.41 as obtained by \cite{Boillet_2018} after fitting their model to the data from \cite{Planck_tsz_2016}. The only frequency dependence of the tSZ power spectra comes from the $f(\nu)$ factors, and this has been divided out in the plot. Aat about $\ell \sim 1000,$ the two-halo term contributes about 2\% to the total power spectrum and the one-halo term dominates. In their Figs. 1 and 2, \cite{Boillet_2018} showed that the tSZ power spectra are very sensitive to the choice of the halo mass function, to the $c-M_h$ relation (where $c$ is the halo concentration parameter), and to cosmological parameters. This means taht several sources of uncertainties arise when the tSZ power spectra are calculated. In Fig. \ref{fig:tsz_f1} we  show the marginalised $y$ power spectrum after subtraction of the foreground residuals as measured by \cite{Planck_tsz_2016}. \cite{Boillet_2018} improved upon the \textit{Planck} analysis by accounting for the trispectrum in the covariance matrix, and they provided new values for the $y$ power spectrum after marginalising over the foreground residuals, which are shown in the figure as well. The best-fit model is consistent with these measurements. At $\ell = 257.5$, using the best-fit parameters mentioned before, we obtain $\ell(\ell + 1) \: C_\ell^\mathrm{tSZ}/2\pi = 0.178$. For this $\ell$, \cite{Planck_tsz_2016} provided the measured $y$ power spectrum ($10^{12}y^2$) value as $0.217\pm0.049,$ which is consistent with our predictions (the best-fit value from their model is 0.203). We placed the error bars on the \textit{Planck} values as the quadrature sum of the statistical and the foreground errors provided by them. At the same time, \cite{Boillet_2018} provided a value of $0.179\pm 0.034$, with a best-fit value of 0.131. This illustrates the range of values that the tSZ power spectra can have based on the different assumptions in the modelling. 

\section{CIB-tSZ cross-correlation} \label{sec:cib-tsz-halo}
As pointed out in the introduction, CIB and tSZ are expected to be correlated to a certain degree. Thus in order to interpret this correlation, we need a cross-correlation model that predicts this signal. In the sections before, we presented the halo models for calculating the CIB and tSZ power spectra, and we use them below to calculate the CIB-tSZ correlation in a consistent halo model formalism.

\subsection{Halo model formalism} \label{subsec:cib-tsz-halo}
The one-halo term provides the correlation between the CIB sources and the tSZ within the same halo. On the other hand, the two-halo term arises from the correlation between the CIB sources in one halo with the tSZ in other halo. This formulation has a very interesting consequence. In an extreme scenario where a very massive halo contributes significantly to the tSZ without star formation, it hence has no CIB sources. In this case, the one-halo term would be zero, but the two-halo term would still contribute provided that there is some overlap in the redshift distribution of the CIB sources and the tSZ haloes. This means that the two-halo term, which dominates on large angular scales, does not depend significantly on the astrophysical processes governing the star formation in massive haloes contributing to the tSZ.  

The  corresponding one- and two-halo terms are given as (for $k = \ell/\chi$)
\begin{align} \label{eq:tsz-cib-1h}
C_{\ell, \nu, \nu'}^{1h} &= \int dz \frac{dV}{dz} \int d\log M_h \frac{d^2N}{d\log M_h d\log V} \: y_\ell \notag \\
& \times \Big[ \big\{ \frac{dj'_{\nu_1, c}}{d\log M_h} + \frac{dj'_{\nu_1, \mathrm{sub}}}{d\log M_h} u(k, M_h, z) \big\} f(\nu_2) \notag \\
& + \big\{ \frac{dj'_{\nu_2, c}}{d\log M_h} + \frac{dj'_{\nu_2, \mathrm{sub}}}{d\log M_h} u(k, M_h, z) \big\} f(\nu_1) \Big] \, ,
\end{align}
and
\begin{align} \label{eq:tsz-cib-2h}
C_{\ell, \nu, \nu'}^{2h} &= \int dz \frac{dV}{dz} P_\mathrm{lin}(k, z) \notag \\
& \times \int d\log M_h \frac{d^2N}{d\log M_h d\log V} \: y_\ell \, b(M_h, z) \notag \\
& \times \int d\log M'_h \frac{d^2N}{d\log M'_h d\log V} \, b(M'_h, z) \notag \\
& \times \Big[ \big\{ \frac{dj'_{\nu_1, c}}{d\log M'_h} + \frac{dj'_{\nu_1, \mathrm{sub}}}{d\log M'_h} u(k, M'_h, z) \big\} f(\nu_2) \notag \\
& + \big\{ \frac{dj'_{\nu_2, c}}{d\log M'_h} + \frac{dj'_{\nu_2, \mathrm{sub}}}{d\log M'_h} u(k, M'_h, z) \big\} f(\nu_1) \Big] \, ,
\end{align}
where 
\begin{equation}
\frac{dj'_{\nu, c(\mathrm{sub})}}{d\log M_h} (M_h, z) = \frac{dj_{\nu, c(\mathrm{sub})}}{d\log M_h} (M_h, z) \frac{a}{\chi^2} \Big(\frac{d^2N}{d\log M_h dV}\Big)^{-1}\, .
\end{equation}
 $M_h$ and $M'_h$  represent the mass of the
two different haloes between which the correlation is computed when the two-halo term is computed. The tSZ spectral variation does not dependent on the redshift or halo mass in the non-relativistic limit, therefore the cross spectra are simpler than they would be otherwise. 

The total \mbox{CIBxtSZ} power is then
\begin{equation}
C_{\ell, \nu, \nu'}^{\mathrm{CIBxtSZ}-\mathrm{tot}} = C_{\ell, \nu, \nu'}^{1h}  + C_{\ell, \nu, \nu'}^{2h} \,.
\end{equation}
When the CIB and tSZ model parameters are knonw, it is straightforward to calculate the CIB-tSZ power spectra and there is no need of additional parameters. This is one of the advantages of the halo model approach over previous studies \citep[e.g.][]{Dunkley_2013, George_2015,  Reichardt_2020, Choi_2020}, where a template-based approach for the CIB-tSZ correlation was used by introducing a parameter $\xi$ to act as a correlation coefficient. This template-based approach is also limited by the fact that $\xi$ might be angular and scale dependent, and fitting for a single parameter $\xi$ might therefore not produce a true picture of the CIB-tSZ correlation. This has indeed been shown to be the case by \cite{Addison_2012}, who found non-negligible variation in $\xi$ with the angular scales and the considered frequencies. 

\subsection{Halo mass definition} \label{subsec:halomassdef}
Our models for CIB and tSZ (and thus CIB$\times$tSZ) rely on the halo mass $M_h$. For the tSZ effect, the pressure profile from Eq.~\ref{eq:pressure_tsz} is in general given in terms of $M_{500}$ , which is defined as the mass contained within a radius in which the mean overdensity is 500 times the background critical density. In contrast, the CIB studies are generally carried out with $M_{200}$ as the definition of the halo mass. We have to be consistent with the definition of the halo mass while considering the CIB and tSZ together. One way of doing this is calculating the CIB terms with $M_{200}$, then converting the $M_{200}$ into corresponding $M_{500}$ for the given redshift range and cosmology using the procedure given in \cite{Komatsu_2001}, and then using this $M_{500}$ to calculate the corresponding tSZ terms. We presented the CIB$\times$tSZ results here, 
we worked with a single definition of $M_{500}$ for the halo masses for the CIB and tSZ terms. We verified the effect of using $M_{500}$ instead of $M_{200}$ on the CIB model parameters, and we found that the values of the CIB parameters slightly change. The most notable change was observed in the value of $\eta_\mathrm{max}$ , which was found to be higher when we used $M_{500}$ instead of $M_{200}$:  because with this definition of the mass, the CIB contribution is calculated very close to the centre of the cluster and does not include the star formation that occurs far away. The model accordingly tries to increase the conversion efficiency of the accreted baryons into stars to produce the same level of CIB as observed. The CIB model parameters given in Tab.~\ref{tab:3halo} were calculated using the commonly used $M_{200}$ definition for the halo masses. 

\begin{figure*}[ht]
\centering
\includegraphics[width=\textwidth]{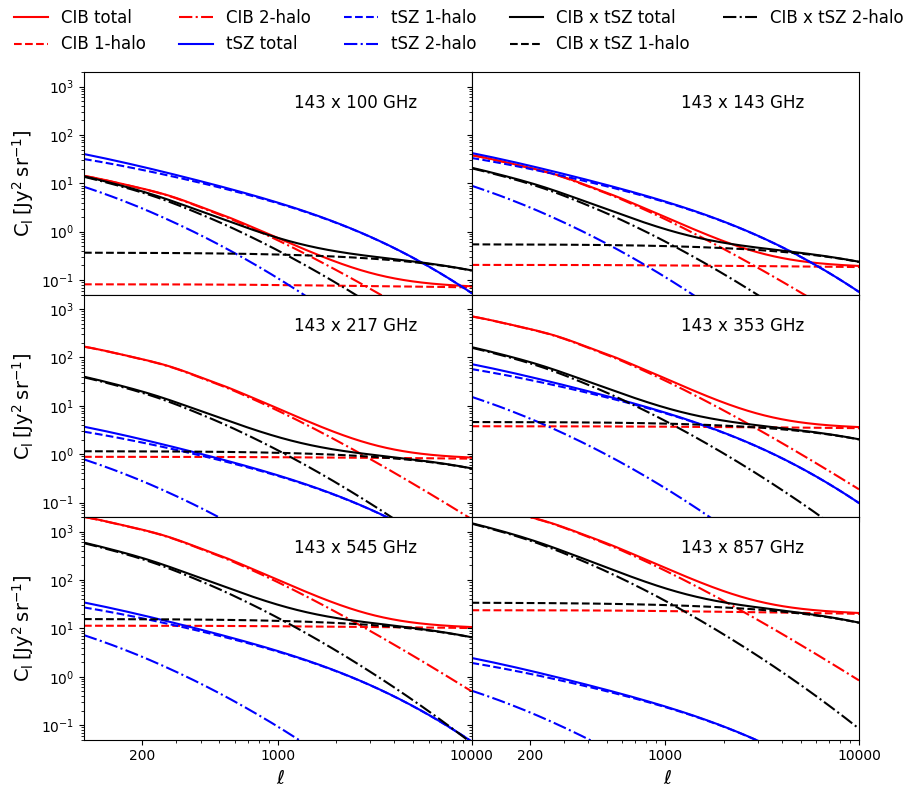}
\centering \caption{Predictions for the different components of the CIB, tSZ, and CIBxtSZ power spectra correlating the \textit{Planck} 143 GHz frequency with other frequencies based on the models given above.}
\label{fig:cib_tsz_cross}
\end{figure*}

\subsection{Predictions for the CIB$\times$tSZ power spectra}\label{ssec:cibxtsz}
Figure~\ref{fig:cib_tsz_cross} shows the predictions for the CIB, tSZ, and CIBxtSZ power spectra correlating the \textit{Planck} 143 GHz frequency with other frequencies based on the models of the CIB, tSZ, and CIBxtSZ given above. We note again that our results were derived using the $M_{500}$ definition for the halo masses, and for the tSZ, the $f(\nu)$ factor was convolved with the \textit{Planck}/HFI frequencies. This provided values of -4.03, -2.79, 0.19, 6.21, 14.46, and 26.34 for 100, 143, 217, 343, 545, and 857 GHz, respectively, as mentioned in Sec.~\ref{sec:tsz_halo}. This means that $f(\nu)$ is negative for $\nu < 217$ GHz and the tSZ power spectra are negative for certain choices of the frequency pairs. In the figure, the tSZ power spectra are negative for a combination of 143\,GHz with all the frequencies above 143 GHz, that is, 217, 353, 545, and 857 GHz (we show their absolute values). Similarly, CIBxtSZ power spectra shown here are negative in all the cases from 100 to 857\,GHz. 
A special case to consider when Eqs.~\ref{eq:tsz-cib-1h} and \ref{eq:tsz-cib-2h} are used is that of the two widely separated frequency channels. Fig.~\ref{fig:cib_tsz_cross} shows that the relative power of the CIB-tSZ to CIB increases with the frequency separation for frequencies $\nu \geq 217$\,GHz. In this case, one of the terms in the square parentheses is generally far smaller than the other and can be neglected because the CIB and tSZ depend differently on frequency. The cross-correlation is then driven by the dominating term. The CIB at higher frequencies traces the haloes at lower redshifts, and this increases the overlap with the tSZ clusters. This results in the increase in the CIB-tSZ power compared to the CIB power spectra as the frequency separation increases. This scenario can provide a good opportunity to constrain the CIB-tSZ cross-power spectra. The tSZ term dominates at lower frequency channels and drives the cross-correlation. The tSZ has a null point at $\nu \approx 217$ GHz. The figure shows this, and the location in which the tSZ power spectra in 143x217 GHz are far smaller than for 143x143 and 143x353 GHz. 

The observational constraints for the tSZ and especially the CIB have improved dramatically since results from \cite{Addison_2012}. This has resulted in an improved understanding and modelling of both the tSZ and the CIB, which we take advantage of. Our predictions for the CIBxtSZ correlation for a given frequency channel pair are therefore not exactly the same as those obtained by \cite{Addison_2012} in terms of their amplitude and/or shape. 

\subsection{Redshift contributions to the power spectra}
\begin{figure*}
\centering
\includegraphics[width=\textwidth]{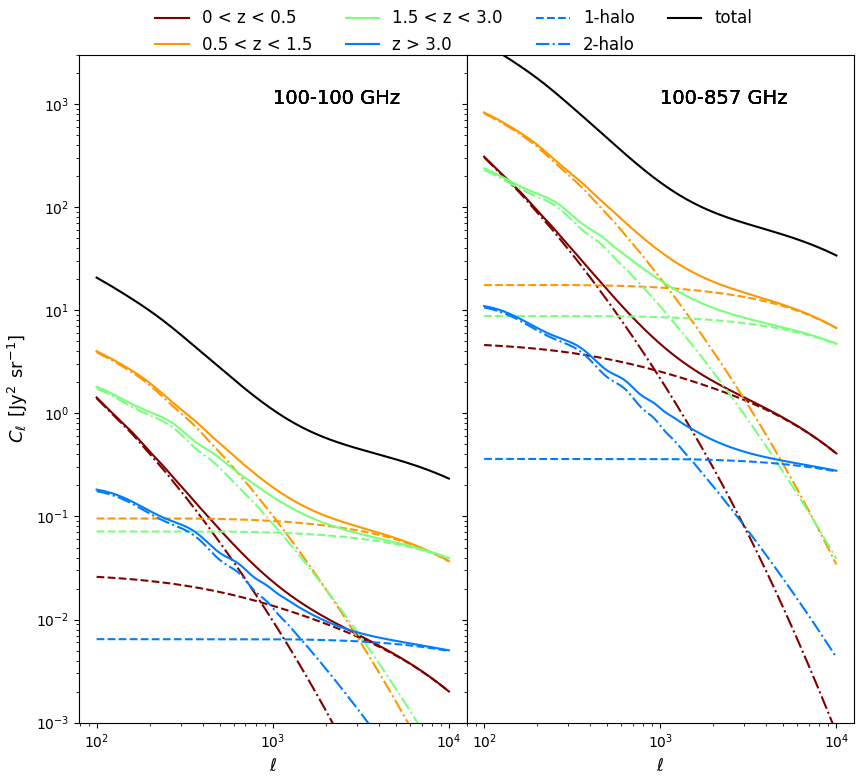}
\centering \caption{Redshift contribution to the power spectra of the CIB-tSZ correlation at for 100-100 GHz channel pair (right panel) and 100-857 GHz channel pair (right panel). All the values shown here are absolute values. }
\label{fig:cibxtsz_reds}
\end{figure*}

Figure~\ref{fig:cibxtsz_reds} shows the contributions to the CIB-tSZ power spectra from different redshift bins. The power spectra are shown for 100-100 GHz and 100-857 GHz channel pairs.  

The clusters hosting hot gas with energetic electrons that cause the tSZ effect reside at relatively low redshifts ($\sim z < 1$). We therefore expect the contribution of the one-halo term of the power spectrum to the total power spectrum to relatively decrease compared to the two-halo term at higher redshifts. This is indeed what we observe in Fig.~\ref{fig:cibxtsz_reds}, where the multipole where the one-halo and two-halo term contribute equally moves slightly to higher values in higher redshift bins, that is, the range in which the two-halo term dominates the one-halo term increases with redshift. This of course depends upon the width of the redshift bins considered and on the pair of frequency channels considered, but the overall trend is the same. 

 Fig.~\ref{fig:cibxtsz_reds} shows that although majority of the tSZ clusters are expected to reside at lower redshifts, the CIB-tSZ contribution coming from the lowest redshift bin ($0 < z < 0.5$) is smaller than or equal to that from other redshift bins. For the 100-100 GHz channel pair, our model predicts that the CIB-tSZ contribution coming from $0 < z < 0.5$ is of the same order as the $z > 3$ bin where we do not expect to find many tSZ clusters, while for the 100-857 GHz channel pair, the contribution from the $0 < z < 0.5$ redshift bin is higher than the $z > 3$ bin. The CIB at lower frequencies traces the galaxies at higher redshift to a certain extent, and vice versa. In the 100-100 GHz case, the CIB contribution mainly comes from galaxies at relatively higher redshift \citep[e.g. see Fig.~4 of][]{Maniyar_2018}, and thus the two-halo term caused by the overlap in the redshift distribution of the CIB sources and the tSZ haloes gives a slightly higher power in the $z > 3$ redshift bin than the $0 < z < 0.5$ bin in the 100-100 GHz case. The case for the 100-857 GHz pair is exactly the reverse: the CIB is mainly fed by galaxies at lower redshifts, thereby increasing the power in the lowest redshift bin. 

Most of the contribution to the CIB-tSZ power spectra appears to stem from the intermediate redshift bins, that is, $0.5 < z < 1.5$ and $1.5 < z < 3.0$. Going back to Fig.~4 of \cite{Maniyar_2018} again, this is expected because most of the CIB power is coming from the dusty star forming galaxies between $0.5 < z < 4$. Thus the CIB part driving the CIB-tSZ correlation has most of its contribution coming from these redshifts. Again, we can see that going from 100-100 GHz pair to 100-857 GHz pair, the contribution of the $0.5 < z < 1.5$ slightly increases in comparison to $1.5 < z < 3$ bin as even lower redshifts are probed by the CIB at 857 GHz than at 100 GHz. 

\subsection{Angular scale dependence of the CIB$\times$tSZ power spectra} \label{ssec:cibxtsz_temp}
Measuring the kSZ power spectrum from the CMB power spectrum is one of the challenging goals of current CMB cosmology data analysis. It is important to measure the kSZ power spectrum because it can help us to understand the reionisation history of the Universe better. On very small angular scales ($\ell > 2000$) in the CMB power spectrum, the kSZ power spectrum starts to show a similar amplitude as other components such as the CIB, tSZ, and their cross-correlation, which act as foregrounds to correctly measure the kSZ. Figures\,5 and 6 from \cite{George_2015} show that the kSZ is degenerate with the tSZ and CIB$\times$tSZ correlation. A consistent modelling of these foregrounds is highly desirable. 

The kSZ has been constrained by \cite{Dunkley_2013}, \cite{George_2015}, \cite{Reichardt_2020}, and \cite{Choi_2020} using the CMB power-spectrum measurements from the Atacama Cosmology Telescope (ACT) or the South Pole Telescope (SPT), while \cite{Planck_cib_tsz_2016} reported a measurement of the CIB$\times$tSZ correlation using \textit{Planck} data. \cite{George_2015} and \cite{Reichardt_2020} calculated the CIB$\times$tSZ power spectra at every step of their analysis using the CIB and tSZ power spectra obtained with templates at this step, whereas \cite{Dunkley_2013} and \cite{Planck_cib_tsz_2016} used the CIB$\times$tSZ template provided by \cite{Addison_2012}. All of them fit for a single scale-independent amplitude parameter for the CIB$\times$tSZ correlation across all the frequency channels. \cite{Addison_2012} used the CIB model from \cite{Xia_2012} to calculate the CIB$\times$tSZ cross-correlation. One of the shortcomings of this model is that it assumes the spectral properties of the CIB, that is, SEDs, luminosity to be independent of the host halo mass. Our understanding of the CIB and its properties has improved significantly in the last few years, and we know that the spectral properties of the CIB are highly dependent on the host halo mass. We therefore compare the CIB$\times$tSZ correlation calculated with our newly developed CIB halo model with that developed by \cite{Addison_2012} using the CIB model from \cite{Xia_2012}. 

\begin{figure}[ht]
\centering
\includegraphics[width=9cm]{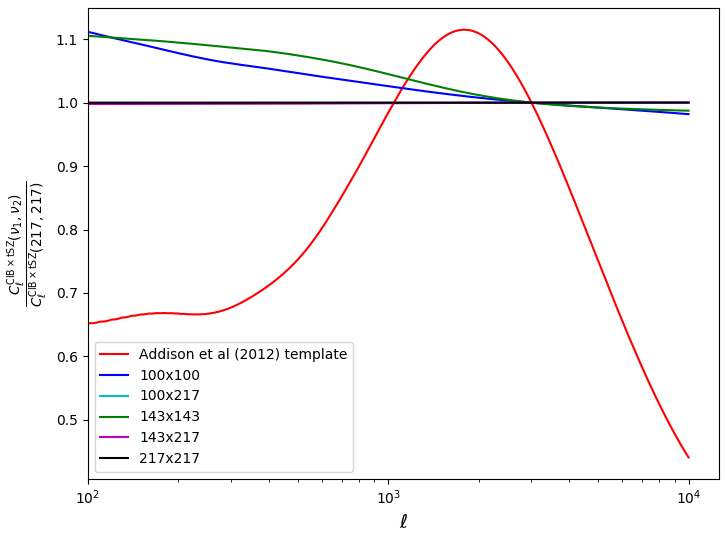}
\centering \caption{Ratio of the CIB$\times$tSZ power spectra at different frequencies and at 217$\times$217 GHz. The red curve shows the ratio with the CIB$\times$tSZ template from \cite{Addison_2012} used in the \cite{Planck_like_2019} likelihood analysis. All the power spectra are normalised such that they have the same value at $\ell = 3000$.}
\label{fig:cib_tsz_temp}
\end{figure}

Figure\,\ref{fig:cib_tsz_temp} shows the ratio of the CIB$\times$tSZ template from \cite{Addison_2012} used in the \cite{Planck_like_2019} likelihood analysis and the corresponding power spectra calculated within our halo model framework at 217$\times$217 GHz in red. The two power spectra were normalised such that they have equal value at $\ell = 3000$. The ratio of the CIB$\times$tSZ template used in the \cite{Planck_like_2019} with our model is different at all scales (e.g. 1.1 at $\ell \sim 2000$ and 0.9 at $\ell \sim 3700$). The results for the kSZ power spectra derived using these templates might therefore be different than if the halo model developed in this paper were used. In Fig.\,\ref{fig:cib_tsz_temp} we also plot the ratio of the CIB$\times$tSZ power spectra for different pairs of \textit{Planck} frequency channels with that at 217$\times$217\,GHz calculated using our halo model, again normalised such that they have the same value at $\ell=3000$. Although at very high multipoles the ratio  does not vary strongly because the CIB at these nearby frequency channels is highly correlated, it does vary by up to 10\% for $\ell < 2000$. \cite{Addison_2012} also showed that for widely separated frequency channels, the CIB$\times$tSZ correlation is different at different multipoles. Fitting for a single amplitude parameter independent of the multipoles for the CIB$\times$tSZ power spectra across all frequency channels is therefore a crude approximation. The best-fit templates used in the previous analysis should be replaced by such physically motivated halo models. This will matter even more for future CMB data that will have a very high signal-to-noise ratio on the small angular scales of interest for the kSZ power spectrum.

\subsection{Correlation coefficient for CIB$\times$tSZ } \label{ssec:cibxtsz_rho}
\begin{figure}[ht]
\centering
\includegraphics[width=9cm]{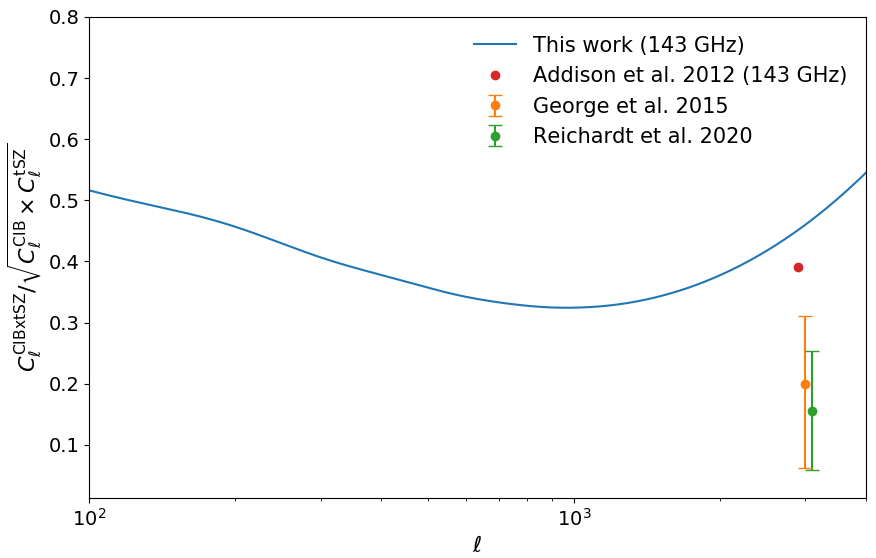}
\centering \caption{Correlation coefficient for the CIB$\times$tSZ cross-correlation calculated with the halo model for the 143 GHz \textit{Planck} channel. We also show the corresponding measurements for the correlation coefficient from \cite{Addison_2012}, \cite{George_2015}, and \cite{Reichardt_2020} for SPT frequency channels at $\ell=3000$. They are slightly offset on the X-axis for clarity.}
\label{fig:cib_tsz_rho}
\end{figure}

We define the correlation coefficient for the CIB and tSZ correlation as 
\begin{equation}
\xi_\ell (\nu) = \frac{C_{\ell, \nu}^\mathrm{CIB x tSZ}}{\sqrt{C_{\ell, \nu}^\mathrm{CIB} \times C_{\ell, \nu}^\mathrm{tSZ}}} \, ,
\label{eq:rho}
\end{equation}
where the correlation coefficient $\xi_\ell (\nu)$ depends on frequency and angular scale. The CIB power spectra $C_{\ell, \nu}^\mathrm{CIB}$ also include the contribution from the shot noise at that particular frequency. In Fig.\,\ref{fig:cib_tsz_rho} we show $\xi_\ell (\nu)$ for 143 GHz \textit{Planck} frequency channel based on the CIB, tSZ, and the CIB$\times$tSZ best-fit models in this paper. We did not fit for the shot noise of the CIB at 143 GHz. We took this value as calculated by the model from \cite{Bethermin_2012}, which is 0.87 $\mathrm{Jy}^2/\mathrm{sr}$. This was added to the one- and two-halo terms at 143 GHz. The CIB$\times$tSZ correlation at 143 GHz is negative, so that we took its absolute value. Based on the halo model formalism in \cite{Addison_2012}, they provide the value of $\xi_\ell$ for cross-correlation of 150 GHz SPT  frequency channel with other frequencies. They also provide the factors to convert these values into those that would be obtained for the corresponding \textit{Planck} 143 GHz channel that we used here. \cite{George_2015} and \cite{Reichardt_2020} measured the value for $\xi_\ell$ at $\ell=3000$ using the 95, 150, and 220 GHz SPT frequency channels. We show these points in Fig.~\ref{fig:cib_tsz_rho}. 

The correlation coefficient $\xi_\ell$ is defined by \cite{Addison_2012}, \cite{George_2015}, and \cite{Reichardt_2020} in a slightly different manner. The values shown in the plot here are twice as high as the values provided by them because of this change in definition. Moreover, as mentioned in Sec.~\ref{subsec:cib-tsz-halo}, \cite{George_2015} and \cite{Reichardt_2020} used a template-based approach to calculate the CIB and the tSZ power spectra. They then used these values to calculate the CIB$\times$tSZ correlation by fitting for the cross-correlation coefficient $\xi$. They therefore obtained a single value of $\xi_\ell$ for the power spectra across all the frequencies, that is, 95, 150, and 220 GHz across all multipoles. These values were additionally calculated using a different telescope (SPT), which corresponds to a different flux cut ($\sim$6.4\,mJy compared to 710\,mJy for \textit{Planck}). 
Although we can compare the $\xi_\ell$ provided by \cite{Addison_2012} with our predictions, it is therefore not a direct comparison with the values from \cite{George_2015} and \cite{Reichardt_2020}. We show these points just to give an idea of the value they derived for $\xi_\ell$. \\
\cite{George_2015} and \cite{Reichardt_2020} reported a value of $0.20\substack{+0.14 \\ -0.11}$ and $0.16\substack{+0.10 \\ -0.10}$ , respectively, at $\ell = 3000$. Our best-fit model gives a value of $\xi_\ell$ of 0.45, which is higher than the value at $\ell = 3000$ reported by \cite{Addison_2012}, which is 0.39. There should be an error bar on $\xi_\ell$ calculated here and by \cite{Addison_2012} because of the uncertainties corresponding to the CIB and the tSZ halo models used. The calculation of these uncertainties is left for future work. 
\section{Conclusions} \label{sec:concl}
One of the main motivations of this work was designing a consistent framework for calculating the CIB, tSZ, and CIB$\times$tSZ power spectra in a halo model setting. For this purpose, we developed a simple and physically motivated halo model of the CIB with only four parameters describing the relationship between the mass of the dark matter haloes and their efficiency to convert the accreted baryons into stars using a lognormal parametrisation. Because previous evidence showed that massive haloes do not contribute significantly at lower redshift to the total star formation budget but are efficient at high redshift, we allowed the width of the lognormal to evolve with redshift. 
We find that the mass of the dark-matter haloes for the highest efficiency is $\log_{10}M_\mathrm{max} = 12.94\substack{+0.02 \\ -0.02} \: M_\odot$, while the maximum efficiency at this mass is $\eta = 0.42\substack{+0.03 \\ -0.02}$. The mass of the highest efficiency found here is slightly on the high side of the range $\log_{10}M = 12.1\substack{+0.50 \\ -0.50}$ M$_{\odot}$ to $12.6\substack{+0.10 \\ -0.10}$ M$_{\odot}$ found by \cite{Viero_2013} and \cite{Planck_cib_2014}, but it agrees well with \cite{Chen_2016} for faint SMGs of $\log_{10}M = 12.7\substack{+0.1 \\ -0.2}$ and $\log_{10}M = 12.77\substack{+0.128 \\ -0.125}$ using the linear clustering model for the CIB anisotropies \citep{Maniyar_2018}.
This agreement is quite motivating considering the simple nature of our model. The model is also able to fit the SFRD measured using the galaxies and is consistent with the SFRD history obtained using the linear clustering model from \cite{Maniyar_2018}. The SFRD constraints derived from the galaxy surveys are added for the first time as priors in the likelihood for the halo model. This helps us achieve the significant result of reproducing the CIB power spectra and the SFRD history at the same time with this model. We also calculated the CIB-CMB lensing cross-correlation using the best-fit value of the CIB parameters and find that it agrees well with the measurements (Fig.~\ref{fig:lens_fit_halo}). While we obtain a decent $\chi^2$ of 113 for 80 \textit{Planck} data points, the $\chi^2$ for \textit{Herschel} power spectra comes is 247 for 102 data points, which is not good. We investigated this discrepancy using different methods and found that the CIB power spectrum measurements are not fully compatible with each other for $\ell>$3000. 

For the tSZ, we followed the approach of \cite{Boillet_2018} and developed a halo model to calculate the one- and two-halo power spectra. Because we kept the parameters for the pressure profile of the gas constant, the power spectra calculations were sped up by tabulating the values of $y_\ell$. The only parameter for the tSZ power spectra is the mass bias $B,$ which takes the difference between the true halo mass and the mass derived assuming the hydrostatic equilibrium, for example, into account. Although we find that the contribution of the two-halo term is not significant $\sim 2\%$ at $\ell \sim 1000$ to the total power spectra, we still considered it in our calculations for a complete analysis and because the two-halo term has a significant contribution to the two-halo term of the CIB$\times$tSZ, which is significant and cannot be ignored. 

Following our two halo models for the CIB and tSZ, we computed the cross-correlation between the CIB and tSZ within a halo model framework. 
In addition to being consistent with the CIB and the tSZ halo models, another advantage of this approach is that we do not require an additional parameter to calculate the correlation. When the CIB and tSZ model parameters are known, it is straightforward to calculate the CIB-tSZ correlation. This is quite advantageous compared with other studies that used the template approach for the power spectra and a fit for the global amplitude. Based on our model and the best fit of the CIB and tSZ parameters, we find that the relative power of the CIB-tSZ correlation with respect to the CIB and tSZ increases with frequency separation of the maps. The two-halo term for the CIB-tSZ correlation is also not negligible with respect to the one-halo term and should be considered in the total power spectrum calculation. This two-halo term arises from the correlation between the tSZ, which is fed from one halo, and the CIB galaxies, which are located in the other halo. Moreover, most of the contribution to the CIB-tSZ power spectra comes from $0.5 < z < 4$ because most of the power of the CIB part that contributes to the CIB-tSZ comes from these redshifts. 

CIB and tSZ act as foregrounds in CMB anisotropy measurements. Especially on small scales, the CIB, tSZ, and CIB$\times$tSZ correlation acts as a hindrance for measuring the kinematic SZ (kSZ) signal, which is the SZ effect caused by the peculiar velocities of the galaxy clusters after reionisation and by the ionised bubbles during the reionisation. In order to measure the kSZ from the CMB power spectrum, it is therefore important to correctly model and remove these foregrounds. Previous analyses that measured the kSZ power spectra have used a template-based approach to model the foregrounds and fit for a single amplitude parameter across all the frequency channels independent of the multipole range considered. We showed that this approach is not sufficient to capture the scale dependence of the CIB$\times$tSZ power spectra, which is moreover model dependent. A similar argument is also applicable to modelling the other foregrounds, that is, the CIB and tSZ as well. The kSZ constraints obtained using the template approach have thus to be revised, replacing the templates by power spectra computed by the physically developed halo models. Because our halo model is effective, it might be used for this purpose. \\

\begin{acknowledgements}
We acknowledge financial support from the "Programme National de Cosmologie and Galaxies" (PNCG) funded by CNRS/INSU-IN2P3-INP, CEA and CNES, France and support from the OCEVU Labex (ANR-11-LABX-0060) and the A*MIDEX project (ANR-11-IDEX-0001-02) funded by the "Investissements d'Avenir" French government program managed by the ANR. GL has received funding from the European Research Council (ERC) under the European Union's Horizon 2020 research and innovation programme (grant agreement No 788212) and funding from the Excellence Initiative of Aix-Marseille University-A*Midex, a French “Investissements d’Avenir” programme". AM warmly thanks Marco Tucci for useful discussions  and inputs on the halo modelling of the CIB power spectra.
\end{acknowledgements}

\bibliographystyle{aa}
\bibliography{bib2}


\appendix\label{app}

\section{The SFRD prior importance} \label{app:sfrd}
Along with providing a good fit to the CIB power spectra, the halo model should be able to reproduce the SFRD history measured by extrapolating the galaxy luminosity functions. For this reason, we used the SFRD measurements from galaxies at different redshifts as priors while fitting for the halo model parameters. The best-fit parameters then result in the SFRD history, as shown in Fig.~\ref{fig:sfrd_halo}, which is consistent with the external measurements. 

\begin{figure}[ht]
\centering
\includegraphics[width=9cm]{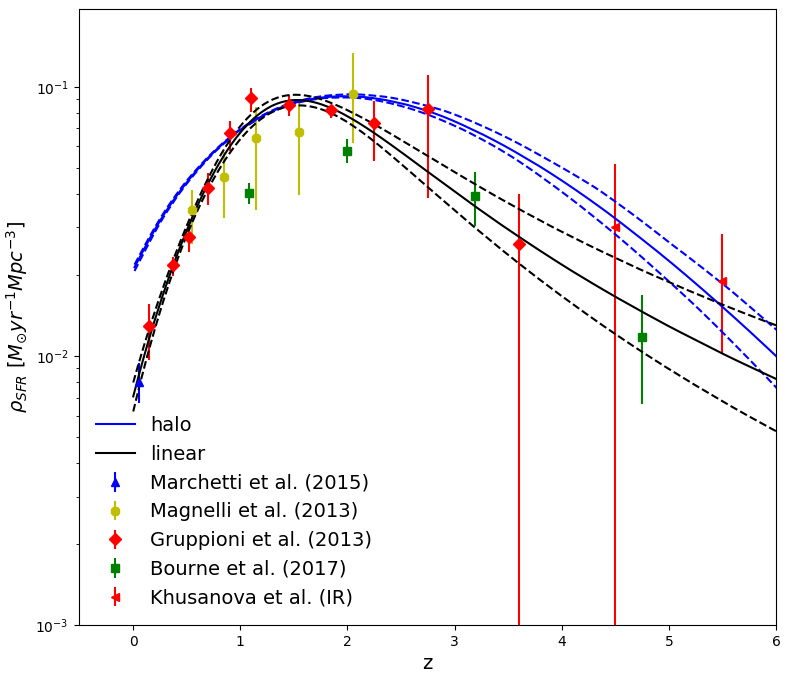}
\centering \caption{Measurements of the SFRD using galaxy surveys \citep{Madau_2014}. The solid black line shows the SFRD as measured by the CIB linear model of \cite{Maniyar_2018}. The solid blue line shows the corresponding constraints when the SFRD values from the galaxies are not considered as priors while performing the fit of the CIB halo model. The dotted black and red lines show the 1$\sigma$ regions.}
\label{fig:sfrd_nosfrdprior}
\end{figure}

Figure~\ref{fig:sfrd_nosfrdprior} shows the results when these external measurements are not considered as priors when the best-fit parameters of the CIB model are defined. In this case, the $\chi^2$ value obtained is 90 for 80 \textit{Planck} data points and 156 for 102 \textit{Herschel} data points.  However, the best-fit value for the calibration factor for 1200 GHz $f_{1200}^\mathrm{cal}$ , which is negatively correlated with the shot-noise at 1200 GHz, is 0.74 against the Gaussian prior set upon it centred at 1.00 with 1 $\sigma$ error bar of 0.05. This $\chi^2$ is much better especially for \textit{Herschel} data than the case when we considered the external SFRD measurements as priors in our likelihood. Although we are able to fit the CIB power spectra much better with these parameters, it is therefore evident that they predict an excessive SFRD, at least at low redshifts. It is therefore quite important to include these measurements as priors in our model, which comes at the expense of a poor $\chi^2$ value. 

\section{Alternate parametrisations} \label{app:moster}


In addition to the lognormal parametrisation considered for $\eta$, we studied several different parametrisations. One was inspired by the approach taken by \cite{Moster_2013}, who connected the galaxy mass with the corresponding host halo mass through a double power-law parametrisation. For our case, we used a double power law to describe the relation between $\eta$ and $M_h$ , which reads
\begin{equation}
\frac{\mathrm{SFR}}{\mathrm{BAR}} = \eta = \eta_\mathrm{max} \Big[ \big(\frac{M_h}{M_\mathrm{max}}\big)^{-\beta} + \big(\frac{M_h}{M_\mathrm{max}}\big)^{\gamma} \Big]^{-1} \, ,
\end{equation}
where $\beta$, and $\gamma$ describe the $\eta$ at the low- and high-mass end, respectively. These two parameters allow us to have asymmetrical distribution around the mass of maximum efficiency. We fit this parametrisation with all the data sets and external constraints as mentioned in Sect.~\ref{sec:priors}. The data used are not sensitive and thus cannot constrain the low-mass end slope $\beta$. Therefore we fixed the value of $\beta$ and let $\gamma$ evolve with redshift (for $z \leq 1.5$) to avoid significant contribution by very massive haloes to the SFR at low redshift, in line with the reasoning presented in Sec.~\ref{ssec:accr_sfr}. Although this parametrisation gave a good fit to the data (even better than the fiducial model using lognormal) and was able to produce a decent SFRD history, the contribution from the haloes above the mass of maximum efficiency at high redshift was unrealistically low. Even though this parametrisation provided a better fit to the data than the fiducial model, we therefore continued with the lognormal because the results were more physical. 

\section{Comparison of the \textit{Planck} and \textit{Herschel} CIB power spectra}\label{app:comparison}
Here we show some figures and plots that compare the measurements of the CIB power spectra from \textit{Planck} and \textit{Herschel} as well as the best-fit values for the halo model obtained under different conditions. The shot-noise values provided here have to be multiplied with the corresponding colour corrections at each frequency to obtain them in the $\nu I_\nu = \mathrm{constant}$ convention.
\begin{figure}[ht]
\centering
\includegraphics[width=9cm]{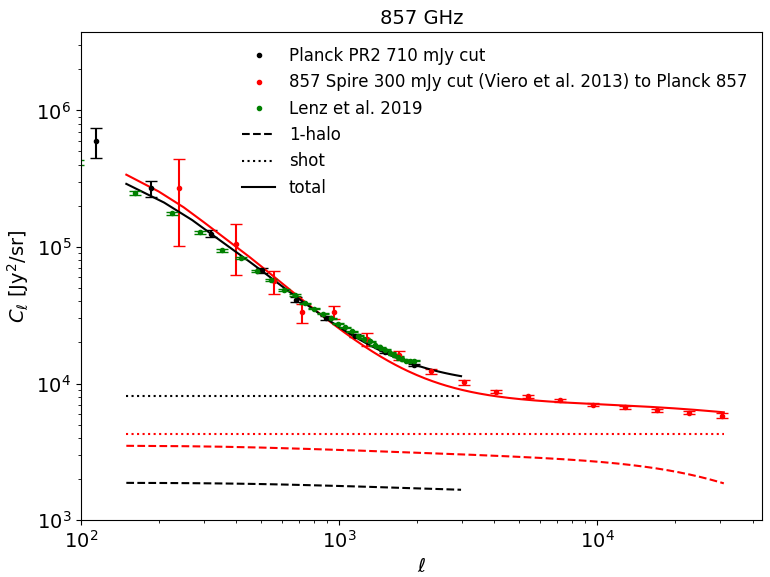}
\centering \caption{Measurements of the CIB power spectra from \cite{Planck_cib_2014}, \cite{Viero_2013}, and \cite{Lenz_2019}. The best-fit models shown in this figure are obtained when \textit{Planck} and \textit{Herschel} data are fit separately.  The SPIRE data have been rescaled to \textit{Planck} data using the cross-calibration factor derived in \cite{Bertincourt_2016} and the factor that allows converting the  measurement through the SPIRE bandpass into a measurement as it would be obtained through the \textit{Planck} bandpass from \cite{Lagache_2019}. Overall, $C_\ell^\mathrm{Planck, 857 GHz} = 1.008 \times C_\ell^\mathrm{SPIRE, 857 GHz}$.
\label{fig:cib_alldata}}
\end{figure}

Figure \ref{fig:cib_alldata} shows the measured \textit{Planck} 545 GHz and SPIRE 600 GHz auto-power spectra in the same plot. On large angular scales, both data sets appear to agree well. However, on small scales, the two data sets differ, which appears to be mostly a result of the inconsistent shot-noise values of \textit{Planck} and \textit{Herschel}. A similar trend appears for the measurements from \cite{Lenz_2019}. A more detailed discussion of this can be found in \cite{Lagache_2019} and is also presented in Sec.~\ref{ssec:res}.

\begin{figure}[ht]
\centering
\includegraphics[width=9cm]{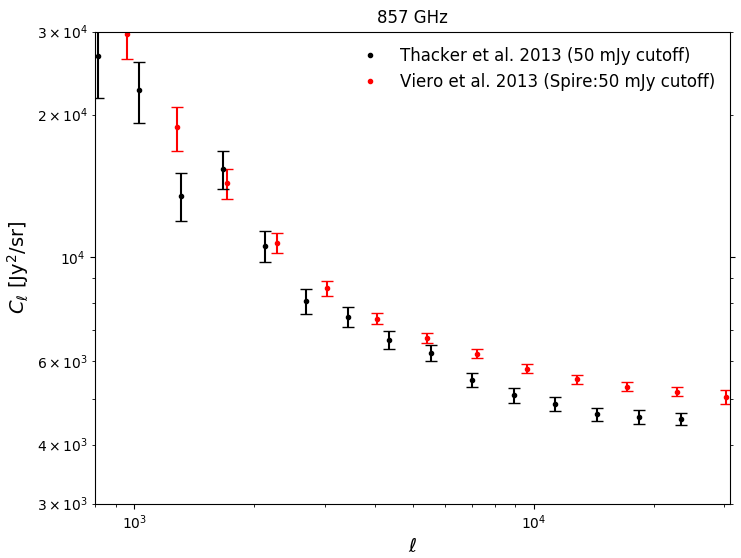}
\centering \caption{Measurements of the CIB power spectra from \cite{Thacker_2013} and \cite{Viero_2013} for a flux cut of 50\,mJy. }
\label{fig:cib_thacker_spire}
\end{figure}

In Fig.~\ref{fig:cib_thacker_spire} we show the comparison of the measured CIB auto-power spectra at 857 GHz for SPIRE by \cite{Thacker_2013} and \cite{Viero_2013} for the same flux cut. Similar to the comparison with \textit{Planck}, the two measurements agree well on the large angular scales. However, they are clearly different on small angular scales, but we expect them to be the same. This shows that there are inconsistencies within the analysis of a single experiment.

In order to determine whether the \textit{Planck} and \textit{Herschel} data can be reconciled within our model, we performed some tests, the results of which are provided in the following tables. Instead of fitting the \textit{Planck} and \textit{Herschel} data together with our model, we only fit for the \textit{Planck} data and present the best-fit values in Tab.~\ref{tab:onlypla}. In this case, instead of assuming a fixed perfect calibration for the \textit{Planck} experiment, we let the calibration factors at all the frequencies vary with a Gaussian prior centred at 1.00 with 1$\sigma$ error of 0.05. The \textit{Planck} $\chi^2$ value improves compared to when we fit for both the \textit{Planck} and \textit{Herschel} data together. The posterior values of calibration parameters are very close to one, and it is therefore a justified assumption to keep the \textit{Planck} calibration fixed while fitting for both the data together.

Similar to the previous test, we then fit our model to \textit{Herschel} data alone. The results are shown in Tab.~\ref{tab:onlyhers}. Unlike the \textit{Planck} data, we obtain a poor $\chi^2$ value in this case. The best-fit values for the shot noise as well as the calibration parameters are consistent with the case when we fit for both the \textit{Planck} and \textit{Herschel} data together. There is some shift in the physical parameters of the halo model, but there is no particular trend as these parameters are correlated with each other.

In order to verify the effect of the cross-calibration between \textit{Planck} and \textit{Herschel}, we performed several tests and show the results in Tab.~\ref{tab:hers+pla+freecal} and \ref{tab:hers+freecal}. In this case, instead of setting tight cross-calibration priors on the \textit{Herschel} calibration parameters, we allowed them to vary with larger Gaussian priors with 1$\sigma$ error bars of 0.05 centred at 1.00. In the first test, we fit for both the \textit{Planck} and \textit{Herschel} data, and in the second case, we fit only for the \textit{Herschel} data. This test indeed provided very good $\chi^2$ values for both \textit{Planck} and \textit{Herschel}. However, it comes at the cost of unrealistically high values of the calibration parameters at 545 and 857 GHz that are completely inconsistent with the very precise cross-calibration measurement between \textit{Planck} and \textit{Herschel} \citep{Bertincourt_2016}. This test therefore does not solve the compatibility problem of the two data sets.

\begin{figure*}[ht]
\centering
\includegraphics[width=\textwidth]{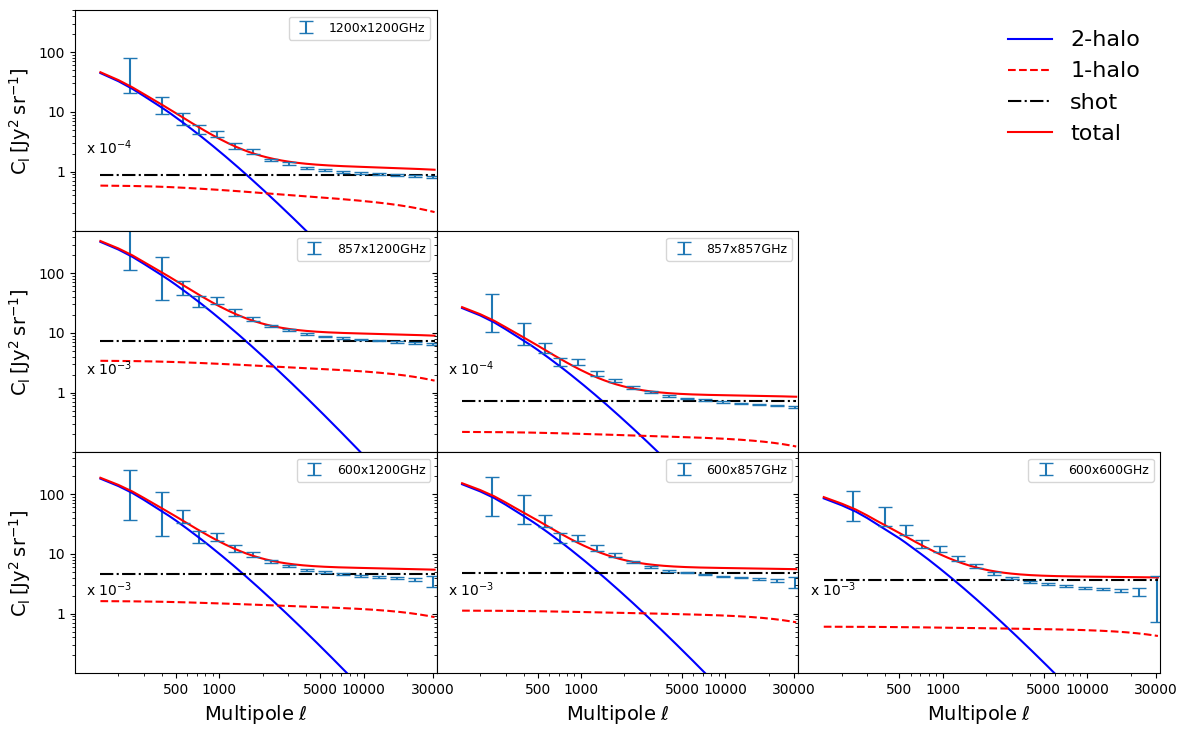}
\centering \caption{Measurements of the CIB auto- and cross-power spectra obtained by \textit{Herschel}/SPIRE \citep{Viero_2013} and the best-fit CIB halo model with its different components when \textit{Herschel}/SPIRE data alone were fit for $\ell<3000$. }
\label{fig:hers+lowell3000}
\end{figure*}

We determined the consistency of the \textit{Planck} and \textit{Herschel} data on the large angular scales by fitting for the \textit{Herschel} data at $\ell<3000$. The results are shown in Tab.~\ref{tab:hers+lowell3000} and Fig.~\ref{fig:hers+lowell3000}. In this case, we obtain a decent $\chi^2$ value, and the shot-noise values are similar to the levels obtained for \textit{Planck;}  this was expected. This tests shows that the \textit{Herschel} and \textit{Planck} data appear to be compatible with each other on large angular scales, but they differ on small scales. 

\begin{table*}
 \centering
\begin{tabular}{ccccc}
\hline
Halo model parameters & $\eta_\mathrm{max}$ & $\log_{10}M_\mathrm{max}$ & $\sigma_{M_{h0}}$ & $\tau$  \\ [6pt]
 $z_c = 1.5$ (fixed) & $0.52\substack{+0.03 \\ -0.02}$ & $12.73\substack{+0.02 \\ -0.02} \: M_\odot$ & $1.24\substack{+0.12 \\ -0.13}$ & $0.82\substack{+0.09 \\ -0.09}$  \\ [6pt]
\hline
HFI shot noise & $\mathrm{SN^{pl}_{217}}$ & $\mathrm{SN^{pl}_{353}}$ & $\mathrm{SN^{pl}_{545}}$ & $\mathrm{SN^{pl}_{857}}$  \\ [6pt]
 & $14\substack{+0.71 \\ -0.71}$ & $357\substack{+9.91 \\ -8.66}$ & $2349\substack{+37.11 \\ -34.16}$ & $7407\substack{+190.48 \\ -179.71}$   \\ [6pt]
\hline
HFI calibration & $f_{217}^\mathrm{cal}$ & $f_{353}^\mathrm{cal}$ & $f_{545}^\mathrm{cal}$ &  $f_{857}^\mathrm{cal}$ \\ [6pt]
 & $1.00\substack{+0.01 \\ -0.01}$ & $1.00\substack{+0.01 \\ -0.01}$ & $0.99\substack{+0.03 \\ -0.03}$ & $1.05\substack{+0.03 \\ -0.03}$ \\ [6pt]
\hline
\end{tabular}
\newline
\centering \caption{Marginalised values of all the model parameters given at a 68 \% confidence level when we fit only for the \textit{Planck} data considering the $f_{\nu}^\mathrm{cal}$ parameters at all \textit{Planck} frequency channels centred at one with error bars of 0.05. We obtain a $\chi^2$ value of 85 for 80 data points. }
\label{tab:onlypla}
\end{table*}




\begin{table*}
 \centering
\begin{tabular}{ccccc}
\hline
Halo model parameters & $\eta_\mathrm{max}$ & $\log_{10}M_\mathrm{max}$ & $\sigma_{M_{h0}}$ & $\tau$  \\ [6pt]
 $z_c = 1.5$ (fixed) & $0.52\substack{+0.04 \\ -0.04}$ & $12.98\substack{+0.02 \\ -0.02} \: M_\odot$ & $1.41\substack{+0.11 \\ -0.10}$ & $0.94\substack{+0.07 \\ -0.07}$  \\ [6pt]
\hline
SPIRE shot noise & $\mathrm{SN^{sp}_{600}}$ &  $\mathrm{SN^{sp}_{857}}$ & $\mathrm{SN^{sp}_{1200}}$ \\ [6pt]
 & $1876\substack{+62.56 \\ -54.50}$ & $4199\substack{+164.88 \\ -148.09}$ & $4602\substack{+399.66 \\ 329.33}$ \\ [6pt]
\hline
HFI/SPIRE cross-calibration & $f_{600}^\mathrm{cal}$ & $f_{857}^\mathrm{cal}$ &  $f_{1200}^\mathrm{cal}$ \\ [6pt]
 & $1.06\substack{+0.01 \\ -0.01}$ & $1.02\substack{+0.01 \\ -0.01}$ & $1.00\substack{+0.03 \\ -0.03}$ \\ [6pt]
\hline
\end{tabular}
\newline
\centering \caption{Marginalised values of all the model parameters given at a 68 \% confidence level when we fit only for the \textit{Herschel} data considering the $f_{\nu}^\mathrm{cal}$ parameters at all \textit{Herschel} frequency channels, as done in our original analysis. We obtain a $\chi^2$ value of 221 for 102 data points. }
\label{tab:onlyhers}
\end{table*}

\begin{table*}
\centering
\begin{tabular}{ccccc}
\hline
Halo model parameters & $\eta_\mathrm{max}$ & $\log_{10}M_\mathrm{max}$ & $\sigma_{M_{h0}}$ & $\tau$  \\ [6pt]
 $z_c = 1.5$ (fixed) & $0.40\substack{+0.03 \\ -0.02}$ & $12.95\substack{+0.02 \\ -0.02} \: M_\odot$ & $1.87\substack{+0.12 \\ -0.13}$ & $1.42\substack{+0.10 \\ -0.10}$  \\ [6pt]
\hline
HFI shot noise & $\mathrm{SN^{pl}_{217}}$ & $\mathrm{SN^{pl}_{353}}$ & $\mathrm{SN^{pl}_{545}}$ & $\mathrm{SN^{pl}_{857}}$  \\ [6pt]
 & $13\substack{+0.71 \\ -0.71}$ & $352\substack{+9.91 \\ -8.66}$ & $2021\substack{+37.11 \\ -34.16}$ & $7279\substack{+190.48 \\ -179.71}$   \\ [6pt]
\hline
SPIRE shot noise & $\mathrm{SN^{sp}_{600}}$ &  $\mathrm{SN^{sp}_{857}}$ & $\mathrm{SN^{sp}_{1200}}$ \\ [6pt]
 & $1150\substack{+62.56 \\ -54.50}$ & $2528\substack{+164.88 \\ -148.09}$ & $4393\substack{+399.66 \\ 329.33}$ \\ [6pt]
\hline
HFI/SPIRE cross-calibration & $f_{600}^\mathrm{cal}$ & $f_{857}^\mathrm{cal}$ &  $f_{1200}^\mathrm{cal}$ \\ [6pt]
 & $1.24\substack{+0.01 \\ -0.01}$ & $1.19\substack{+0.01 \\ -0.01}$ & $1.04\substack{+0.03 \\ -0.03}$ \\ [6pt]
\hline
\end{tabular}
\newline
\centering \caption{Marginalised values of all the model parameters given at a 68 \% confidence level when we fit for both the \textit{Planck} and \textit{Herschel} data considering the $f_{\nu}^\mathrm{cal}$ parameters at all \textit{Herschel} frequency channels centred at one with error bars of 0.05. We obtain a $\chi^2$ value of 105 for 80 \textit{Planck} data points and 127 for 102 \textit{Herschel} data points with unrealistically high $f_{\nu}^\mathrm{cal}$ values. }
\label{tab:hers+pla+freecal}
\end{table*}

\begin{table*}
 \centering
\begin{tabular}{ccccc}
\hline
Halo model parameters & $\eta_\mathrm{max}$ & $\log_{10}M_\mathrm{max}$ & $\sigma_{M_{h0}}$ & $\tau$  \\ [6pt]
 $z_c = 1.5$ (fixed) & $0.43\substack{+0.04 \\ -0.04}$ & $12.95\substack{+0.02 \\ -0.02} \: M_\odot$ & $1.60\substack{+0.11 \\ -0.10}$ & $1.06\substack{+0.07 \\ -0.07}$  \\ [6pt]
\hline
SPIRE shot noise & $\mathrm{SN^{sp}_{600}}$ &  $\mathrm{SN^{sp}_{857}}$ & $\mathrm{SN^{sp}_{1200}}$ \\ [6pt]
 & $1137\substack{+62.56 \\ -54.50}$ & $2502\substack{+164.88 \\ -148.09}$ & $4389\substack{+399.66 \\ 329.33}$ \\ [6pt]
\hline
SPIRE calibration & $f_{600}^\mathrm{cal}$ & $f_{857}^\mathrm{cal}$ &  $f_{1200}^\mathrm{cal}$ \\ [6pt]
 & $1.27\substack{+0.01 \\ -0.01}$ & $1.21\substack{+0.01 \\ -0.01}$ & $1.05\substack{+0.03 \\ -0.03}$ \\ [6pt]
\hline
\end{tabular}
\newline
\centering \caption{Marginalised values of all the model parameters given at a 68 \% confidence level when we fit only for the \textit{Herschel} data considering the $f_{\nu}^\mathrm{cal}$ parameters at all \textit{Herschel} frequency channels centred at one with error bars of 0.05. We obtain a $\chi^2$ value of 138 for 102 \textit{Herschel} data points with unrealistically high $f_{\nu}^\mathrm{cal}$ values. }
\label{tab:hers+freecal}
\end{table*}

\begin{table*}
\centering
\begin{tabular}{ccccc}
\hline
Halo model parameters & $\eta_\mathrm{max}$ & $\log_{10}M_\mathrm{max}$ & $\sigma_{M_{h0}}$ & $\tau$  \\ [6pt]
$z_c = 1.5$ (fixed) & $0.52\substack{+0.09 \\ -0.09}$ & $12.86\substack{+0.02 \\ -0.02} \: M_\odot$ & $1.22\substack{+0.19 \\ -0.16}$ & $0.81\substack{+0.13 \\ -0.11}$  \\ [6pt]
\hline
SPIRE shot noise & $\mathrm{SN^{sp}_{600}}$ &  $\mathrm{SN^{sp}_{857}}$ & $\mathrm{SN^{sp}_{1200}}$ \\ [6pt]
& $3441\substack{+62.56 \\ -54.50}$ & $7204\substack{+164.88 \\ -148.09}$ & $8326\substack{+399.66 \\ 329.33}$ \\ [6pt]
\hline
SPIRE calibration & $f_{600}^\mathrm{cal}$ & $f_{857}^\mathrm{cal}$ &  $f_{1200}^\mathrm{cal}$ \\ [6pt]
 & $1.05\substack{+0.01 \\ -0.01}$ & $1.01\substack{+0.01 \\ -0.01}$ & $1.03\substack{+0.03 \\ -0.03}$ \\ [6pt]
\hline
\end{tabular}
\newline
\centering \caption{Marginalised values of all the model parameters given at a 68 \% confidence level when we fit only for the \textit{Herschel} data upto $\ell<3000$ considering the $f_{\nu}^\mathrm{cal}$ parameters at all \textit{Herschel} frequency channels as done in our original approach. We obtain a $\chi^2$ value of 36 for 42 \textit{Herschel} data points. }
\label{tab:hers+lowell3000}
\end{table*}

\end{document}